\documentclass[fleqn,10pt]{wlscirep}
\usepackage[utf8]{inputenc}
\usepackage[T1]{fontenc}
\usepackage{bm}
\usepackage{siunitx}
\usepackage{soul}
\usepackage[normalem]{ulem}
\usepackage[skins,theorems]{tcolorbox}
\tcbset{highlight math style={enhanced,
  colframe=red,colback=white,arc=0pt,boxrule=1pt}}
\usepackage{amsmath,hyperref}
\usepackage{amsfonts,amssymb}
\usepackage{lineno}
\usepackage{ulem}
\hypersetup{
     colorlinks=true,
     linkcolor=blue,
     filecolor=blue,
     citecolor = blue,      
     urlcolor=blue,
     }
\usepackage{tcolorbox}
\usepackage{tabularx}
\usepackage{array}
\usepackage{colortbl,ulem}
\tcbuselibrary{skins}

\newcommand\redout{\bgroup\markoverwith
{\textcolor{red}{\rule[.5ex]{2pt}{0.4pt}}}\ULon}

\newcolumntype{Y}{>{\raggedleft\arraybackslash}X}

\tcbset{tab1/.style={fonttitle=\bfseries\large,fontupper=\normalsize\sffamily,
colback=yellow!10!white,colframe=red!75!black,colbacktitle=yellow!40!white,
coltitle=black,center title,freelance,frame code={
\foreach \n in {north east,north west,south east,south west}
{\path [fill=red!75!black] (interior.\n) circle (3mm); };},}}

\tcbset{tab2/.style={enhanced,fonttitle=\bfseries,fontupper=\normalsize\sffamily,
colback=yellow!10!white,colframe=red!50!black,colbacktitle=yellow!40!white,
coltitle=black,center title}}

\title{Atomistic mechanisms of viscosity in 2D liquid-like fluids}
\author[1]{Dong Huang}
\author[1]{Shaoyu Lu}
\author[1]{Chen Liang}
\author[2$\ddagger$]{Matteo Baggioli}
\author[1$\ddagger$]{Yan Feng}
\affil[1]{Institute of Plasma Physics and Technology, Jiangsu Key Laboratory of Frontier Material Physics and Devices, School of Physical Science and Technology, Soochow University, Suzhou 215006, China}
\affil[2]{Wilczek Quantum Center, School of Physics and Astronomy, Shanghai Jiao Tong University, Shanghai 200240, China \& Shanghai Research Center for Quantum Sciences, Shanghai 201315, China}
\affil[$\ddagger$]{ \color{blue}b.matteo@sjtu.edu.cn\color{black}; \color{blue}fengyan@suda.edu.cn\color{black}\vspace{0.2cm}}

\begin{abstract}
\textbf{Shear viscosity plays a fundamental role in fluid dynamics from heavy-ion collisions to biological processes. Still, its microscopic mechanisms at the individual particle kinetic level remain a subject of ongoing research, specially in dense systems. In this work, we systematically investigate the shear viscosity ($\eta$) of two-dimensional (2D) simple fluids using computer simulations of Lennard-Jones, Yukawa, and one-component plasma systems. By combining Frenkel's liquid description, consisting of solid-like quasi-harmonic vibrations interrupted by thermally activated hops, with the concept of lifetime of local atomic connectivity $\tau_{LC}$, we find a surprisingly simple formula for the kinematic viscosity that is solely determined by $\tau_{LC}$ and the average kinetic particle speed $\bar{v}_p$. The derived analytical expression provides a direct link between macroscopic and microscopic dynamics, which shows excellent agreement with the simulation data in the dense liquid-like regime in all the 2D fluids considered. Moreover, it is discovered that, $\tau_{LC}$ in 2D fluids is universally determined by the effective potential difference between the first peak and valley of the pair correlation function, implying a direct connection between macroscopic shear transport and microscopic structure. Finally, we demonstrate that the characteristic length scale $l_p= \bar{v}_p \tau_{LC}$, which governs the macroscopic shear viscosity, aligns with the elastic length-scale that defines the propagation limit of collective shear waves in liquids. These findings establish that shear viscosity in 2D fluids arises from the diffusive transport of average particle momentum across the elastic length scale. Moreover, they highlight that shear dynamics are fundamentally governed by localized configurational excitations within the atomic connectivity network.}
\end{abstract}

\begin{document}
\flushbottom
\maketitle
\begin{figure}
   \centering
   \includegraphics[width=0.85\linewidth]{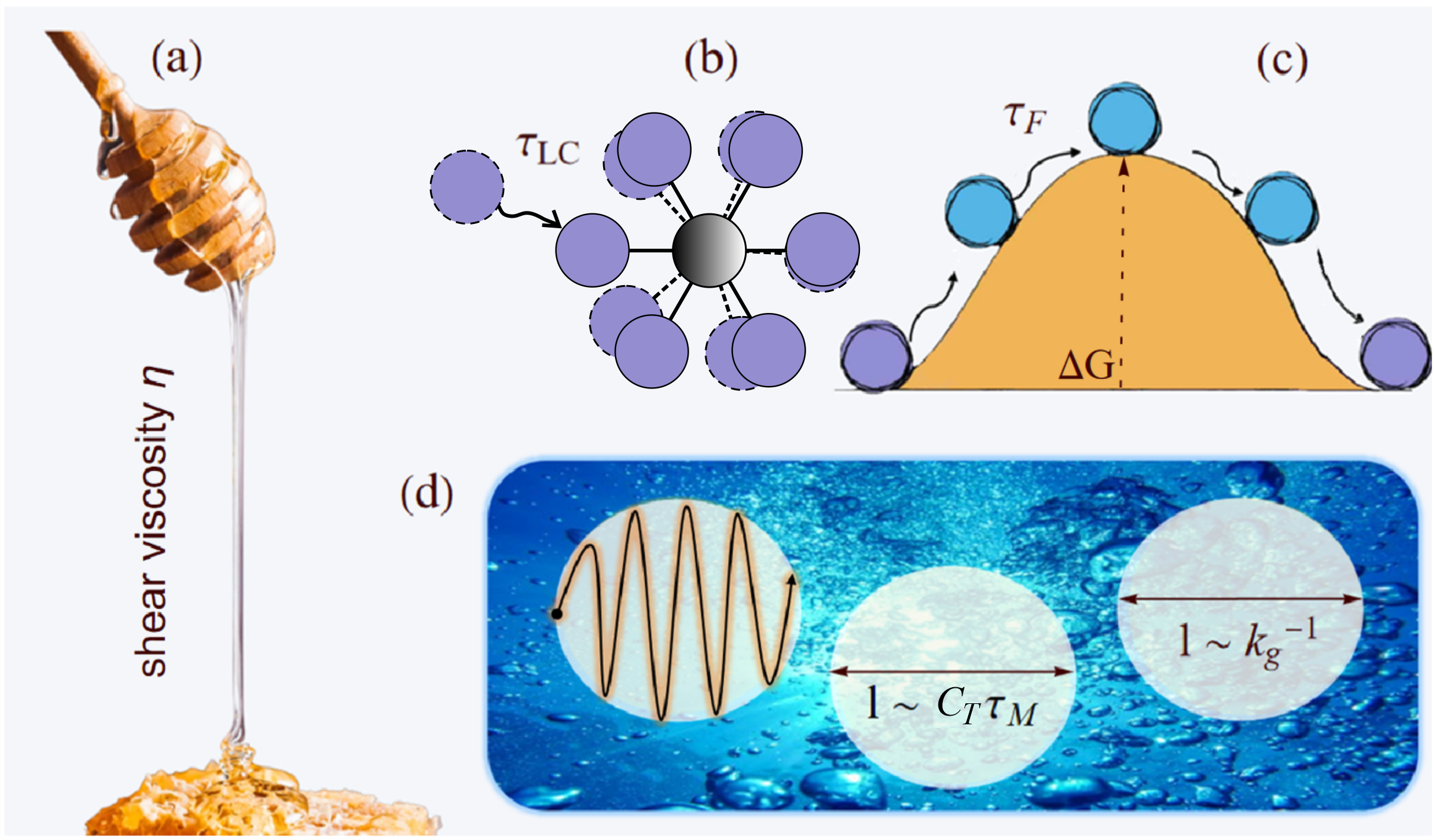}
   \caption{\textbf{Viscosity, particle motion, and collective shear dynamics in liquids:} \textbf{(a)} The shear viscosity $\eta$ determines the macroscopic resistance to shear flow in fluids. \textbf{(b)} A local configurational excitation consisting in losing or gaining one neighbor. $\tau_{LC}$, the lifetime of local connectivity, is the average timescale associated to this microscopic process. \textbf{(c)} Structural rearrangements in liquids are governed by localized events in which one or few particles hop a potential barrier ($\Delta G$), as assumed in Eyring and Frenkel theories of liquid viscosity. This activated process happens with an averaged rate $\tau_F^{-1}$, where $\tau_F$ is the microscopic Frenkel time. \textbf{(d)} Collective shear waves in liquids propagate only up to a length-scale $l\sim 1/k_g$, with $k_g$ the wave-vector gap in their dispersion. According to Maxwell viscoelasticity theory, $k_g \sim 1/(C_T \tau_M)$ where $C_T$ is the high-frequency speed of propagation for shear waves and $\tau_M$ is the collective Maxwell relaxation time.}
  \label{fig:1}
\end{figure}

\thispagestyle{empty}
As first described by Newton in his 1687 \textit{Principia}, the shear viscosity ($\eta$) characterizes the momentum flux transport in fluids, defined as the ratio of shear stress to shear rate~\cite{landau1987fluid,hansen2013theory} (see Fig.~\ref{fig:1}\color{blue}(a)\color{black}). Newton's law of viscosity provides a macroscopic framework that can be translated into an operative definition using the Green-Kubo formalism \cite{Evans_Morriss_2008} (see \cite{Evans01071978} for an example of its applications), nevertheless, it does not offer any physical insights about the microscopic and kinetic origin of viscosity.

In gases, a microscopic description of viscosity has been already achieved more than $200$ years ago with the formulation of kinetic theory, where momentum transport is accomplished via molecular collisions~\cite{loeb2004kinetic}. This led to the known relation of $\eta \propto \rho \bar{v}_p l_{\text{mfp}}$ and the typical temperature dependence of $\eta \propto \sqrt{T}$ in the dilute gas regime, where $\rho$ is the mass density, $\bar{v}_p$ is the mean speed of particles, and $l_{\text{mfp}}$ is the mean free path. Here, $l_{\text{mfp}}$ is the average distance between two consecutive collisions, which is independent of temperature. However, in liquids, molecular dynamics are inherently collective and cooperative, rendering traditional kinetic theory inapplicable as potential energy contributions play a significant role. This leads to a completely different and much stronger temperature dependence $\eta \propto \exp(\Delta G/k_B T)$, indicating that viscosity in liquids is governed by the activation energy barrier for molecular rearrangements $\Delta G$, while not by thermal collisions.

In line with the original picture of liquid dynamics proposed by Frenkel \cite{Frenkel1946}, consisting of solid-like quasi-harmonic vibrations in the local basins interrupted by thermally activated hops over potential barriers, numerous studies have attempted a description of liquid viscosity based on activated-rate theory of chemical reactions. Eyring theory ~\cite{eyring1935activated} is the most famous example of this sort, predicting that $\eta= A \exp \left({\Delta G} / {k_B T}\right)$. Here, $A$ is an undetermined pre-factor and $\Delta G$ is the atomic hopping potential barrier (see Fig.~\ref{fig:1}\color{blue}(c)\color{black}). The Eyring equation for viscosity correctly captures the experimentally observed exponential decrease in liquid viscosity with temperature~\cite{osti_5437529}. However, it remains a semi-empirical theory, since both $A$ and $\Delta G$ are left undetermined. 

On the other hand, in Frenkel's description of liquids dynamics \cite{Frenkel1946}, the potential barrier $\Delta G$ appearing in Eyring's formula is interpreted as the energy necessary to re-arrange the nearest-neighbor cage. The typical timescale of these configurational re-arrangements was formally defined by Egami \cite{PhysRevLett.110.205504} using the concept of lifetime of local atomic connectivity $\tau_{LC}$. By definition, $\tau_{LC}$ (see Fig.~\ref{fig:1}\color{blue}(b)\color{black}) refers to the average time in which one particle loses or gains one neighbor within the surrounding cage. It was realized that above the Arrhenius temperature this timescale coincides with the collective Maxwell relaxation timescale $\tau_M=\eta/G_\infty$ (with $G_\infty$ the instantaneous shear modulus), providing a microscopic interpretation of structural relaxation and viscosity in high-temperature liquids. This observation suggests a connection between viscosity and local configurational excitations in liquids that was further discussed in \cite{PhysRevE.98.022604,10.1063/1.4789306}, revealing surprising similarities with the Drude model for electric transport in metals \cite{PhysRevE.98.063005}. By looking at the atomic scale stress correlation function, the relevant length-scale for viscosity was also related to the range of propagation of shear waves \cite{10.1063/1.4789306} (see Fig.~\ref{fig:1}\color{blue}(d)\color{black}), providing another link to a modern version of Frenkel's ideas known as $k$-gap theory \cite{Trachenko_2016,BAGGIOLI20201}, and also to the dual model of liquid viscosity \cite{peluso2024viscosityliquidsdualmodel}. 

It is important to note that a microscopic and exact expression for viscosity in terms of liquid structure already exists. In 1947, Born and Green achieved this by formulating a relation based on the radial distribution function \cite{doi:10.1098/rspa.1947.0088}. Since then, the Born-Green formalism has seen significant development and has been extended to account for strain-rate-dependent viscosity as well (see, for example, \cite{EVANS1980321,due}). However, the Born-Green theory requires solving a complex integro-differential equation (see Chapters 5 and 6 in \cite{doi:10.1098/rspa.1947.0088}) and does not provide direct insight into the underlying dynamics governing liquid viscosity.

Another microscopic formula for liquid viscosity was recently suggested based on viscoelastic non-affine motion \cite{PhysRevE.108.044101} but the identification of the relevant degrees of freedom was not properly clarified. A follow-up analysis \cite{huang2024microscopicoriginliquidviscosity} showed that, within that framework, only unstable localized normal modes contribute to viscosity with a possible crossover to a stable mode dominated regime at low temperatures. The results of \cite{huang2024microscopicoriginliquidviscosity} align with the idea of Egami that viscosity in liquids is governed by localized events. Interestingly, in water, the timescale relevant for shear viscosity was proven to correlate with the connectivity of the fluctuating hydrogen bond network and the evolution of the first and second nearest neighbors \cite{doi:10.1126/sciadv.1603079,D0CP01560A}.

More recently, the microscopic origin of viscosity for 2D Yukawa fluids was revisited under the view of Egami's idea and discovered to coincide with the momentum transfer process of losing/gaining nearest neighbors for individual particles~\cite{PhysRevResearch.4.033064}, leading to the phenomenological expression $\eta=nm \bar{v}_p^2 \tau_{L C}$ (with $\bar{v}_p = \sqrt{2 k_B T / m}$ the averaged particle speed and $\rho=m n$, where $n$ is the number density). Nevertheless, the validity of this expression was verified only in one specific system and was not derived using any physical argument.

In summary, although viscosity is a fundamental property of fluids, a successful and
 universal theory based on particle-level motion, akin to kinetic theory for dilute gases,
 is still missing, particularly in the dense regime. In this work, we propose a simple,
 approximate theory of viscosity that captures the main mechanisms of stress relaxation
 and momentum transport in the dense fluid state. We present two analytical and closed-form formulae for the viscosity based on (I) single particle motion and local configurational excitations and (II) pure structural information, encoded in the short range behavior of the pair correlation function. We demonstrate the validity of these expressions using extensive simulations in 2D Lennard-Jones (L-J), Yukawa, and one-component plasma (OCP) systems, proving the universality of our findings. Finally, we demonstrate that the length-scale relevant for liquid viscosity aligns with the propagation length of collective shear waves, bridging particle level motion to collective dynamics in liquids, and unifying the previous theoretical frameworks for liquid viscosity. 
\section*{Fluid viscosity from microscopic particle motion}
Following Frenkel's liquid description~\cite{Frenkel1946}, self diffusion can be regarded as a hopping process between potential minima for individual atoms/molecules, where the average distance between two potential minima is assumed to be $\xi$, while the average hopping period is given by the Frenkel time $\tau_F$. By assuming simple random walk motion for the liquid constituents, the diffusion constant $D$ can be written in terms of these two parameters as $D=\xi^2/(4 \tau_F)$. At the same time, the mobility $\alpha$ in a 2D liquid is expressed using Stokes law as $\alpha=1/(4\pi\eta)$, and the Einstein's relation implies $D=\alpha k_B T$. By combining these expressions, we can obtain a simple formula for the shear viscosity of 2D liquids,
\begin{equation}
    \tcbhighmath[fuzzy halo=1mm with blue!50!white,arc=2pt,
  boxrule=0pt,frame hidden]{\eta = \frac {k_{B} T \tau_F} { \pi \xi^2}.}\label{ff}
\end{equation}
This equation is 
a straightforward 2D adaptation of the original Frenkel's liquid formula for 3D systems (see Ref. \cite{Frenkel1946}). Despite the elegance of this formula, its usefulness is questionable since nor $\tau_F$ or $\xi$ are explicitly defined and hence they cannot be estimated from simulation or experimental data.

In order to overcome this problem, we first identify the Frenkel time with the lifetime of local atomic connectivity, $\tau_F = \tau_{LC}$. This is reasonable since particle hopping in the potential landscape corresponds to local structural rearrangements described in real space by changes in the short-range topology.
We then introduce the mean squared particle velocity, defined by $\bar{v}_p^2 = 2k_B T / m$, and rewrite the expression as:
\begin{equation}
\label{frenkeleq2D}
    \eta = \frac{1}{2\pi \xi^2} \, m \bar{v}_p^2 \tau_{LC}.
\end{equation}
To proceed further, we use that the length-scale $\xi$ is proportional to the mean interparticle distance, introducing a dimensionless constant $\delta$ of order one:
\begin{equation}
    \xi = \frac{\delta}{\sqrt{n}},\label{def}
\end{equation}
where $n$ is the number density. Notably, in first approximation, $\xi$ does not depend on temperature. Substituting this into the expression for $\eta$, we obtain:
\begin{equation}
    \eta = \frac{1}{2\pi \delta^2} \, \rho \bar{v}_p^2 \tau_{LC},
\end{equation}
where $\rho = m n$ is the mass density. Finally, we define the prefactor $\lambda \equiv 1 / (2\pi \delta^2)$ to arrive at the compact form:
\begin{equation}
\tcbhighmath[fuzzy halo=1mm with blue!50!white,arc=2pt,
  boxrule=0pt,frame hidden]{  \eta=\lambda \rho \bar{v}_p^2 \tau_{LC} ,}\quad \label{maineq}
\end{equation}
with $\lambda$ being an unknown constant of order unity, encapsulating the uncertainty in the precise relation between $\xi$ and the interparticle spacing.
Eq.~\eqref{maineq} is our final theoretical prediction for the viscosity that involves only one unknown constant (in temperature and density) parameter $\lambda$.

In summary, we have shown analytically that Eq.~\eqref{maineq} is equivalent to Frenkel's expression for the liquid viscosity Eq.~\eqref{ff} upon identifying the Frenkel time with $\tau_{LC}$. We notice that, despite the simplifying assumptions, we are not able to derive the value of $\lambda$. Nevertheless, by invoking phonon dynamics and the identity $\bar{v}_p^2 \tau_{LC} \approx C_T^2 \tau_M$, that has been directly verified in 2D Yukawa liquid-like fluids \cite{PhysRevResearch.4.033064}, where $C_T$ is the transverse sound speed, one can speculate that $\lambda \approx 1$.

We also emphasize that expression for viscosity, Eq. \eqref{maineq}, relies on only two parameters: the average particle speed $\bar{v}_p$ and the lifetime of local connectivity $\tau_{LC}$. The first is a straightforward thermodynamic quantity, readily obtained from the system’s temperature and particle mass. The second is easily accessible in simulations (e.g., \cite{PhysRevLett.110.205504,yu2025understandingflowbehaviorssupercooled}). Moreover, it is experimentally measurable in more macroscopic systems, such as colloidal and granular fluids (see, for instance, \cite{jiang2024experimental}).

Before proceeding, we notice that the the application of Stokes' law to atomistic fluids, and particularly to 2D systems, might be questionable. For instance, in two dimensions, a logarithmic dependence on system size is generally expected (see, e.g., Eq.~(9) in \cite{10.1063/1.4834696}). Despite our use of an inverse proportionality between mobility and viscosity may be a simplifying assumption, our numerical results (see below) indicate that any corrections to this relation are small within the regime of interest. Along these lines, it is important to emphasize that the very concept of viscosity may be ill-defined in 2D systems due to the so-called hydrodynamic long-time tails \cite{PhysRevA.1.18,PhysRevLett.25.1254}. While a comprehensive  of this issue lies beyond the scope of the present work, we have conducted an initial analysis in the Supplementary Material (SM). Our findings show no clear evidence of a  $1/t$ long-time tail in the shear stress autocorrelation function, nor any significant system size dependence in the viscosity obtained via the Green-Kubo formalism. These results are consistent with previous studies \cite{PhysRevE.52.6123,PhysRevLett.93.155004}, which also found no indication of hydrodynamic long-time tails in the shear stress autocorrelation function in comparable systems. In summary, based on our current analysis, viscosity appears to be a well-defined transport coefficient in the 2D systems studied here. Finally, we emphasize that long-time tails are expected to influence both the definition of the diffusion constant $D$ and the validity of the Stokes–Einstein relation. Although these quantities do not enter directly into our proposed formula, Eq.~\eqref{maineq}, it is important to investigate their role further in future work.

To validate Eq.~\eqref{maineq}, we perform numerical simulations of 2D L-J, Yukawa, and OCP systems. In 2D L-J systems \cite{1981PhyA..106..226B}, the interparticle interaction $\phi (r) = 4 \epsilon\left[(\sigma / r)^{12}-(\sigma / r)^6 \right]$ consists of both repulsive and attractive terms, where $\epsilon$ and $\sigma$ are the energy and distance parameters. In 2D Yukawa systems, the interaction between particles is a screened Coulomb repulsion $\phi (r)=Q^{2} \exp \left(-r / \lambda_{D}\right) / 4 \pi \epsilon_{0} r$, where $\lambda_{D}$ is the Debye length and $Q$ is the particle charge. In 2D OCP systems, the interaction is the classical Coulomb repulsion $\phi (r)=Q^{2} / 4 \pi \epsilon_{0} r$. For our simulated 2D systems under various conditions, the reduced temperature $T / T_m$ is specified from $\ge 1.1$ to 20 at most. Since the concept of melting in 2D systems is subtle, in the Supplementary Material we clarify how the ``melting temperature'' $T_m$ (used with a slight abuse of terminology) is defined for the various systems considered. We emphasize that $T_m$ serves only as a reference scale for normalizing the temperature axis and does not affect the physical conclusions of our work.

Besides the temperature $T$, we also vary the number density $n$ of 2D L-J systems and the screening parameter $\kappa = a/\lambda_{D}$ of 2D Yukawa systems, where $a$ is the Wigner-Seitz radius~\cite{Melzer:1996,RevModPhys1353,PhysRevLett145003,PhysRevLett065003,PhysRevLett235001}. All simulation details are provided in the Methods section. Our simulations provide a rather large sample of 2D fluids with remarkably different particle interactions under various conditions, allowing us to test in detail the universality of our findings. For more details about the phase diagram of these systems and the regions explored in our analysis, we refer to the Supplementary Material. We nevertheless anticipate that, unless stated otherwise, we always work in the liquid-like fluid phase.

\begin{figure}
    \centering
    \includegraphics[width=\linewidth]{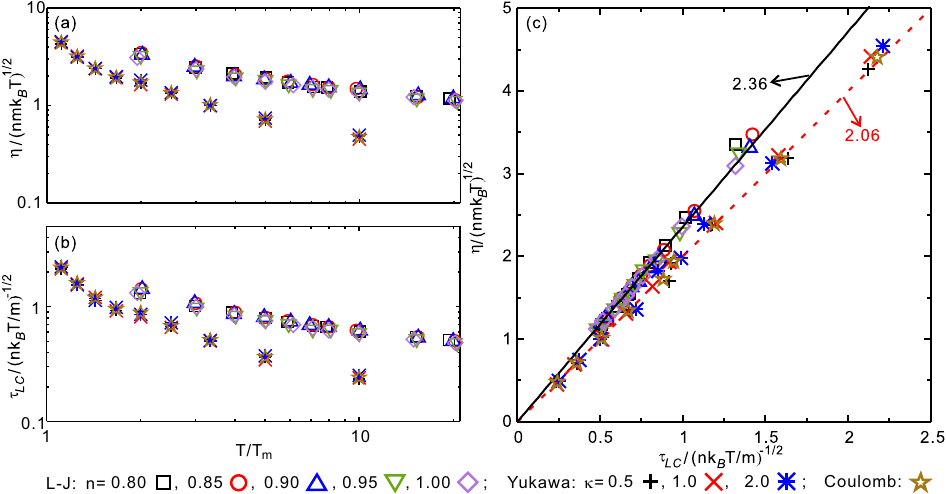}
    \caption{\textbf{ Microscopic origin of viscosity as diffusive transport of average particle momentum:} \textbf{(a)} Dimensionless viscosity as a function of reduced temperature. \textbf{(b)} Dimensionless local connectivity time as a function of reduced temperature. \textbf{(c)} Test of the universal formula for viscosity proposed in Eq.~\eqref{maineq} of the main text.}
    \label{fig:2}
\end{figure}

Our numerical results obtained using the Green-Kubo formalism are presented in Fig.~\ref{fig:2}\color{blue}(a)\color{black}. Following Ref.~\cite{Rosenfeld:2001}, we present the obtained viscosity in a dimensionless form $\eta / \left({nmk}_B T \right)^{1 / 2}$. Interestingly, both the 2D L-J data and the 2D Yukawa and 2D OCP ones collapse into two universal curves as a function of the reduced temperature $T / T_m$. For all systems, the shear viscosity in the liquid phase decreases with temperature, as expected. 

To reveal the fundamental origin of viscosity at the particle level motion, we calculate the lifetime of the local atomic connectivity $\tau_{L C}$~\cite{PhysRevLett.110.205504, Ashwin:2015} for the simulated 2D liquids by tracking the neighbors of each particle (see Methods), as presented in Fig.~\ref{fig:2}\color{blue}(b)\color{black}. By employing the same dimensionless form, $\tau_{LC}/\left(n k_B T / m \right)^{-1/2}$, all the data collapse into two universal curves as well. We notice the discrepancy between these two curves, as in the case of viscosity in Fig.~\ref{fig:2}\color{blue}(a)\color{black}, probably due to the attractive contribution to the potential that is present in the 2D L-J systems but absent in the 2D Yukawa and 2D OCP fluids. 

The similarity in the shape of the normalized viscosity and the normalized local connectivity time is striking, suggesting a direct relation between these two quantities. In order to investigate this point, we rewrite Eq.~\eqref{maineq} in dimensionless units \cite{Rosenfeld:2001},
\begin{equation}
    \frac{\eta}{\left({nmk}_B T \right)^{1 / 2}}= 2 \lambda \frac{\tau_{LC}}{\left(n k_B T / m \right)^{-1/2}}.\label{eq2}
\end{equation}
This identity can now be directly tested using the simulation data presented in Figs.~\ref{fig:2}(a) and~\ref{fig:2}(b). In Fig.~\ref{fig:2}(c), we test directly our proposed expression Eq.~\eqref{eq2} for 2D L-J, Yukawa and OCP fluids. Our numerical results confirm the validity of Eq.~\eqref{maineq} with $\lambda \approx 1.03$ for 2D Yukawa and OCP fluids, and $\lambda \approx 1.18$ for 2D L-J fluid. As already anticipated, these values are surprisingly close to $\lambda=1$. 

By using the values of $\lambda$ obtained from simulations, we can estimate the parameter $\delta$ in Eq.~\eqref{def} using $\lambda \equiv 1 / (2\pi \delta^2)$. We find $\delta \approx 0.39$ for 2D Yukawa and OCP fluids, and $\delta \approx 0.37$ for the 2D Lennard-Jones (L-J) fluid. This corresponds to $\xi \approx 0.69\,a$ for 2D Yukawa and OCP fluids, and $\xi \approx 0.66\,a$ for the 2D L-J fluid, where $a$ is the Wigner-Seitz radius. These results confirm that, to a first approximation, $\xi$ coincides with the Wigner-Seitz radius and is thus proportional to the average interparticle distance. This finding will be further supported by the structural analysis presented below.

In summary, our analysis supports the physical idea that, in 2D simple liquids, the momentum transport process responsible for the macroscopic shear viscosity does originate from losing/gaining neighbors at the individual particle level. Since $\eta/\rho$ in liquids controls the diffusive transport of transverse collective momentum, our results also indicate that this macroscopic dynamical process is associated to a characteristic microscopic time-scale $\tau_{LC}$ and a microscopic length-scale $l_p= \bar{v}_p \tau_{LC}$. We also emphasize that Eq.~\eqref{maineq} establishes a direct link between macroscopic dynamics and microscopic particle motion, providing a fundamental understanding of shear viscosity in liquids at all scales.

We note that the dynamics of the atomic connectivity network underlying the definition of $\tau_{LC}$ merit further attention. An extended analysis is provided in the Supplementary Material (SM), while a more comprehensive study is left for future work.

\section*{Regime of validity of the proposed formula}
\begin{figure}[ht]
    \centering
    \includegraphics{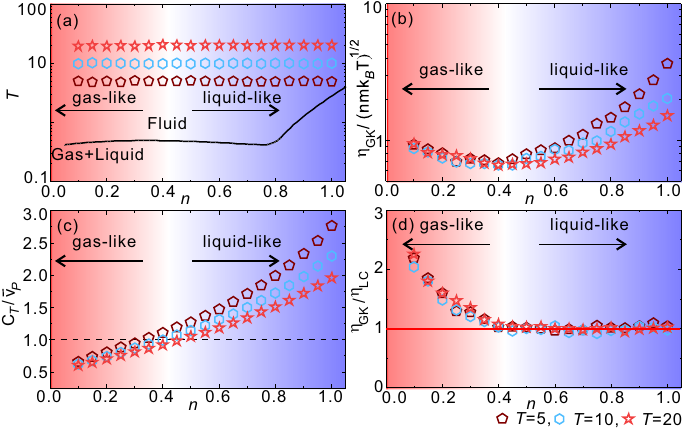}
    \caption{\textbf{(a)} Studied 2D L-J system at constant temperature. The colored symbols indicate the constant-temperature scans in the fluid phase. The background colors show the dynamical crossover (white) between the dilute gas-like phase (red) and the dense liquid-like fluid phase (blue). \textbf{(b)} Normalized viscosity as a function of the density at constant temperature. \textbf{(c)} Ratio of the high-frequency transverse speed of sound $C_T$ to the average particle speed $\bar{v}_p$ as a function of density. \textbf{(d)} Ratio between the viscosity obtained numerically using the Green-Kubo formalism ($\eta_{GK}$) and our theoretical formula Eq.~\eqref{maineq} ($\eta_{LC}$) as a function of density $n$. The same background color scheme is used in all panels.}
    \label{fig:rev}
\end{figure}

In Fig.~\ref{fig:2}, we presented numerical evidence supporting the validity of our viscosity formula, Eq.~\eqref{maineq}. However, it is essential to more rigorously delineate the regime in which our theoretical model holds. The validity of our theoretical model, Eq.~\eqref{maineq}, has been already clearly demonstrated in the liquid-like regime of 2D Yukawa fluids in~\cite{PhysRevResearch.4.033064}. Here, we extend the study of the regime of validity of our viscosity formula to 2D L-J fluids. To this end, we conducted a complementary analysis using the L-J system by scanning across the density axis $n$, from the dilute limit ($n\approx 0.1$) to the dense regime ($n\approx 1$), while keeping the temperature fixed above the critical point, ensuring we remain in the fluid phase \cite{Massimo:2020}. 

A detailed representation of the regimes of exploration within the 2D L-J phase diagram is provided in Fig.~\ref{fig:rev}(a). The background red and blue colors represent respectively the gas-like dilute regime and the dense liquid-like fluid phase. The two phases are separated by a dynamical crossover, that emerges roughly around $n_c \approx 0.4$ for the temperature values considered. This dynamical crossover can be distinguished using several physical observables.

Figure~\ref{fig:rev}(b) displays the normalized viscosity as a function of density, revealing a minimum around $n=n_c$, and confirming the boundary between two regimes with distinct dynamical behaviors and mechanisms of viscosity. Support for this separation comes also from Fig.~\ref{fig:rev}(c), where we plot the ratio of the high-frequency transverse velocity $C_T$ (determined by the instantaneous, or high-frequency, shear modulus, and controlling the propagation of shear waves in the high-frequency regime, $\omega \tau \ll 1$) to the average particle velocity $\bar{v}_p$. This ratio transitions from less than one below $n_c$ to greater than one above, consistent with a dynamical crossover from gas-like to liquid-like behavior (see, e.g., Fig.~1 in \cite{PhysRevResearch.4.033064}). We also note that $n_c$ lies close to the critical density of the L-J fluid.

We then move to testing the validity of our theoretical model for the viscosity. Fig.~\ref{fig:rev}(d) shows the ratio between the viscosity computed via the Green-Kubo formalism ($\eta_{GK}$) and our theoretical prediction ($\eta_{LC}$) from Eq.~\eqref{maineq}. Above the crossover density $n_c$ (white background region), our model yields an excellent estimate of the viscosity, with the ratio $\eta_{GK}/\eta_{LC} \approx 1$ (horizontal red line). In contrast, for $n<n_c$, our formula underestimates the viscosity and fails to capture its density dependence.

So far, we have not identified an exact limiting case for our formula, analogous to the dilute (or high-temperature) limit in kinetic theory. Nevertheless, we have verified that within the liquid-like regime, the averaged error of our theoretical estimates consistently remains between $0.09\%$-$11.5\%$, with an average error of $3.91\%$, for all these three studied 2D fluids when compared to the numerical data.

In conclusion, our viscosity formula, Eq.~\eqref{maineq}, accurately describes the liquid-like regime and breaks down only in the dilute gas-like phase. While this limitation may appear restrictive, it is important to recall that the microscopic origin of viscosity in the gas-like regime is well captured by kinetic theory and its extensions, such as Enskog theory and the hard-sphere fluid model. In this context, our framework complements these established approaches by offering a predictive tool for viscosity and its dependence on temperature and density in the dense liquid-like fluid regime.

\color{black}
\section*{Connecting microscopic particle motion with collective shear dynamics in fluids}
We now take a step back and reconsider the collective shear dynamics in liquid under Maxwell's perspective \cite{maxwell1867iv}. Combining Maxwell approach with Navier-Stokes equations, it has been shown \cite{Trachenko_2016,BAGGIOLI20201} that the dynamics of collective shear waves in liquids are described by the following telegrapher equation:
\begin{equation}
    \omega^2+i \omega/\tau_M= C_T^2 k^2.
\end{equation}
This equation implies that the real part of the shear wave dispersion relation presents a gap in momentum space,
\begin{equation}
    \mathrm{Re}(\omega)=C_T \sqrt{k^2-k_g^2},\qquad k_g=\frac{1}{2 C_T \tau_M},
\end{equation}
where the expression of $k_g$ relies on Maxwell viscoelasticity theory (see \cite{PhysRevE.105.024602} for one alternative interpretation). This $k$-gap feature has been confirmed in many simulation works (\textit{e.g.}, \cite{PhysRevLett.118.215502}) and also in a few experimental setups \cite{PhysRevLett.97.115001,jiang2024experimental}. Most importantly, this result implies that the propagation of collective shear waves in liquids is confined up to an elastic length-scale given by $l\equiv 1/k_g$. In other words, one could construct an idealized model of a liquid as composed of elastic patches of size $l$ in which the dynamics are solid-like and mainly composed of quasi-harmonic oscillations localized at the bottom of the potential \cite{BAGGIOLI20201} (see Fig.~\ref{fig:1}\color{blue}(d)\color{black}). Beyond this elastic length-scale, elastic forces get screened, shear stresses are not supported anymore and the dynamics become liquid-like, \textit{i.e.} dominated by shear diffusive transport rather than coherent wave-like excitations as in solids. Moreover, within Maxwell's theory, the average size of these solid-like regions shrinks with temperature, as a direct consequence of $\tau_M$ decreasing rapidly with $T$. 

Following this idea, one could ask whether this collective elastic length-scale bears any relation to the particle-level length-scale governing viscosity, $l_p= \bar{v}_p \tau_{LC}$, connecting somehow the macroscopic Maxwell view with the particle level Frenkel's description. We notice that the microscopic scale $l_p$ might be associated in Frenkel's picture of liquid dynamics (see Fig.~\ref{fig:1}\color{blue}(a)\color{black}) to the average length $\xi$ made by one particle hopping across potential barriers that is directly related to the viscosity of the system (see Eq.~\eqref{ff}). However, the two length scales have significantly different temperature dependence. $\xi$ is in first approximation temperature independent, while $l_p$ decreases by increasing temperature. It would be interesting to conduct a more in depth analysis about the relation, if any, of these two scales.

\begin{figure}
    \centering
    \includegraphics[width=\linewidth]{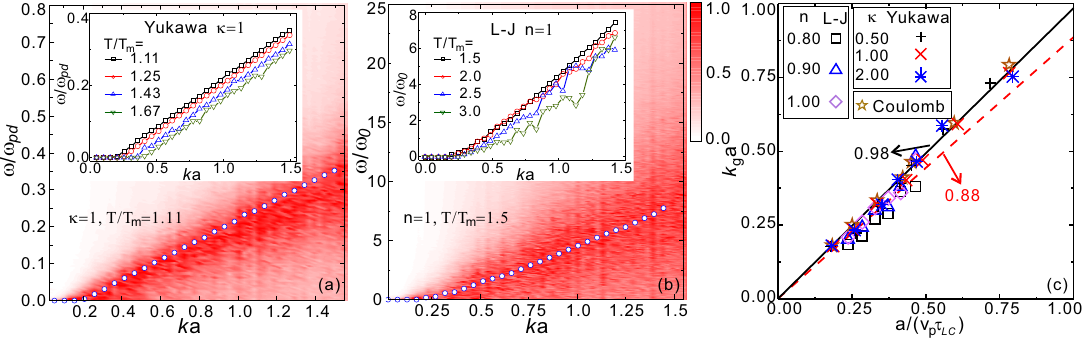}
    \caption{\textbf{Bridging collective shear dynamics to particle-level motion:} \textbf{(a)} Spectra of transverse modes in a 2D Yukawa liquid-like fluid with $\kappa=1$, and the corresponding dispersion relation marked as dots. Frequencies are normalized by the nominal dusty plasma frequency $\omega_{pd}\equiv \left(Q^2/2\pi m \epsilon_0 m a^3\right)^{1/2}$ \cite{PhysRevLett.92.065001}, while wave-vectors are normalized using the Wigner-Seitz radius $a$. The obtained dispersion relations under different reduced temperatures are presented in the inset. \textbf{(b)} Same analysis for 2D L-J liquid-like fluids with particle density $n=1$ at different reduced temperatures $T/T_m$. \textbf{(c)} Universal linear relation between the dimensionless cutoff wave-vector $k_g$ and the inverse dimensionless microscopic length-scale $l_p=\bar{v}_p \tau_{LC}$. The slopes of the two fitting lines are $\approx 0.98$ and 0.88, respectively.}
    \label{fig:3}
\end{figure}

In Fig.~\ref{fig:3}\color{blue}(a) \color{black} we present the numerical results for the dispersion relation of collective shear waves of the 2D Yukawa liquid-like fluids with $\kappa=1$. The presence of a cutoff wave-vector $k_g$ is evident and its size grows with temperature, as expected. By tracking the position at which $\mathrm{Re}(\omega) \rightarrow 0$, we are able to derive the temperature dependence of $k_g$. A similar analysis has been performed for the 2D L-J liquid-like fluids and the corresponding results are presented in Fig.~\ref{fig:3}\color{blue}(b)\color{black}.

Finally, in Fig.~\ref{fig:3}\color{blue}(c)\color{black}, we plot the dimensionless cutoff wave-vector $k_g a$ as a function of the dimensionless inverse length-scale $a/l_p$, with $l_p=\bar{v}_p \tau_{LC}$. For all the systems considered, we find a universal linear relation:
\begin{equation}
\tcbhighmath[fuzzy halo=1mm with blue!50!white,arc=2pt,
  boxrule=0pt,frame hidden]{ k_g = \beta l_p^{-1}}\quad ,\label{wow}
\end{equation}
where $\beta$ is a constant of order one: $\beta\approx 0.88$ for 2D L-J systems and $\beta \approx 0.98$ for 2D Yukawa and OCP systems. 

This result implies a direct proportionality between the elastic length-scale $l \sim k_g^{-1}$, relevant for collective shear dynamics, and the microscopic length-scale $l_p=\bar{v}_p \tau_{LC}$ that governs the macroscopic shear viscosity through Eq.~\eqref{maineq}. It also suggests that the propagation of collective shear waves in liquids is hindered by local configurational excitations that drive structural rearrangements within the nearest neighbor cage.
Furthermore, it is important to stress that, despite a lot of discussions in the past \cite{Trachenko_2016,BAGGIOLI20201}, a formal connection between Frenkel's ideas and $k$-gap theory has never been achieved before. In fact, despite $k$-gap theory was strongly motivated by Frenkel's intuition, the relevant timescale has been always identified with the collective Maxwell relaxation time that, in first approximation, bears no connection with the microscopic Frenkel's time. Eq.~\eqref{wow} provides the missing link between collective shear dynamics, as envisaged by $k$-gap theory, and microscopic particle hops \`a la Frenkel, formalized using the concept of local connectivity time proposed by Egami. Motivated by this observation, we proposed that the length-scale associated to the $k$-gap should not be associated to the Maxwell length-scale $l_m=C_T \tau_M$ but rather to the particle length-scale $l_p$, in better alignment with Frenkel's initial proposal.

\section*{Structural model
 of viscosity}
So far, we have succeeded in connecting the local connectivity time with the macroscopic shear viscosity and showed that the length-scale associated to it aligns (up to an order one constant) with the propagation length of collective shear waves in liquids. In Frenkel's liquid description \cite{Frenkel1946}, fluidity and viscosity arise from particles' hopping over potential barriers, as depicted in Fig.~\ref{fig:1}(c). The average hopping time $\tau_F$ can be expressed as $\tau_F = \tau_0 \exp ({\Delta G} / {k_B T})$, where $\Delta G$ is the potential energy barrier, while $\tau_0$ is the corresponding time in the limit of very high temperatures.

From a structural point of view, a particle hopping over a single potential barrier corresponds to a re-arrangement of its cage or its neighbors. We have therefore advanced the idea that the single particle Frenkel time should be taken to coincide with $\tau_{LC}$. This idea is also supported by the validity of Eq.~\eqref{wow} that has been directly verified in Fig.~\ref{fig:3}. Following this hypothesis, the energy barrier $\Delta G$ should correspond to the energy for one particle hopping outside the cage formed by its neighboring particles. As a result, the pair correlation function $g(r)$ should encode the information about $\Delta G$.

\begin{figure}[htb]
    \centering
    \includegraphics[width=\linewidth]{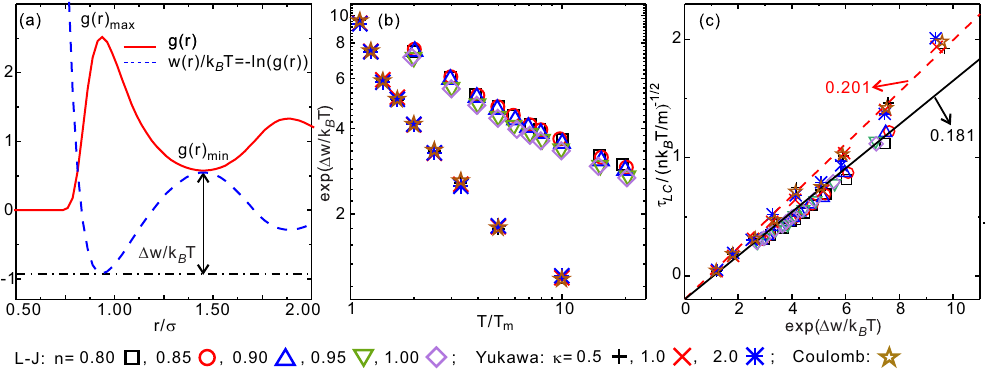}
    \caption{\textbf{Structural definition of the local connectivity time:} \textbf{(a)} Calculated pair-correlation function $g(r)$ of a 2D L-J liquid (red line) and the corresponding effective potential $w(r)/k_B T$ (blue dashed line). The vertical black arrows indicate the potential difference between the first maximum and first minimum that is identified with the potential barrier $\Delta G$ in Frenkel's description, Eq.~\eqref{pot}. \textbf{(b)} The temperature dependence of the potential factor $\exp\left(\Delta w /k_B T\right)$ as a function of the reduced temperature $T/T_m$ for the various systems studied. \textbf{(c)} The universal linear relation between the dimensionless local connectivity time $\tau_{LC}/\left(n k_B T / m \right)^{-1 / 2}$ and $\exp\left(\Delta w/k_B T\right)$ for all systems considered.}
    \label{fig:4}
\end{figure}

In Fig.~\ref{fig:4}\color{blue}(a)\color{black}, we plot the calculated $g(r)$ and the corresponding effective potential $w(r) / k_{B} T = - \ln(g(r))$~\cite{hansen2013theory} for a typical 2D L-J liquid with $n = 1$ and $T/T_m = 5$. We propose that the energy barrier $\Delta G$ governing the hopping of individual particles in the Frenkel description of liquids is given by
\begin{equation}
\Delta G\equiv \Delta w= k_B T \ln \left[ {g(r)_{\rm \max }} / {g(r)_{{\rm \min }}}\right] ,\label{pot}
\end{equation}
where $\textit{max}$ and $\textit{min}$ correspond respectively to the position of the first maximum and first minimum in $g(r)$. The magnitude of $\Delta G$ in Eq.~\eqref{pot} is represented with the vertical black arrows in Fig.~\ref{fig:4}\color{blue}(a)\color{black}.

We then calculate $\exp({\Delta w} / {k_B T})={g(r)_{\rm \max }} / {g(r)_{{\rm \min }}}$ for different 2D liquids under various conditions and present these results as a function of the reduced temperature in Fig.~\ref{fig:4}\color{blue}(b)\color{black}. These obtained data points for $\exp({\Delta w} / {k_B T})$ collapse into two universal curves and present similar variation trends as those for $\eta$ and $\tau_{L C}$ in Fig.~\ref{fig:2}, clearly indicating the strong correlations between these three physical quantities.

To further elucidate this connection, in Fig.~\ref{fig:4}\color{blue}(c) \color{black} we plot the dimensionless local connectivity time as a function of $\exp \left({\Delta w / k_B T}\right)$. We find that these quantities present a robust linear relation independently of the thermodynamic conditions, \textit{i.e.} the value of $T/T_m$. This suggests a simple and striking relation between the local connectivity time and the short-range structural properties of liquids that can be formalized as
\begin{equation}
\tcbhighmath[fuzzy halo=1mm with blue!50!white,arc=2pt,
  boxrule=0pt,frame hidden]{\tau_{L C}=\tau_0 \exp \left(\frac{\Delta w}{k_B T}\right) = \tau_0 \frac{g(r)_{\max }}{g(r)_{{\min }}}}
\quad.\label{vivo}
\end{equation}
Additionally, our numerical analysis points to a simple and natural definition for the timescale $\tau_0$. In particular, we find that $\tau_0 =  \gamma \left(n k_B T / m \right)^{-1 / 2}$, where $\gamma$ is just the slope of the linear fitting in Fig.~\ref{fig:4}(c), i.e., $\gamma \approx 0.181$ for 2D L-J liquid-like fluids while $\gamma \approx 0.201$ for 2D Yukawa and OCP liquid-like fluids. Importantly, despite $\tau_0$ scales with temperature as $1/\sqrt{T}$, mirroring the temperature dependence of the collision time in kinetic theory, its physical interpretation is totally different and cannot be rationalized within ideal kinetic theory. We notice that the empirical expression obtained for $\tau_0$ can be further simplified to relate this timescale to other fundamental quantities. In particular, by using the relation between the density $n$ and the Wigner–Seitz radius $a$, and by assuming an approximate value of $\gamma$ extracted from simulation data, one can infer that $\tau_0 \approx a / (2 \bar{v}_p)$ and $\xi \approx \sqrt{2} \, \bar{v}_p \tau_0$. Interestingly, these findings are consistent with Frenkel's interpretation (see page 200 in \cite{Frenkel1946}), which states that $\tau_0$ is roughly the time required to travel a distance $\xi$ at the thermal velocity $v_{\text{th}}$.

The correction coming from $g(r)_{\max }/g(r)_{{\min }}$ in Eq.~\eqref{vivo} takes into account the short-range and mid-range correlations that become important in the liquid state upon decreasing temperature. This term indeed vanishes if the pair correlation function loses its first peak and first valley, as expected in the ideal gas state. It is immediate to verify that the parameter $\gamma$ is related to $\lambda$ in Eq.~\eqref{maineq} via $\lambda = {1} / ({4 \pi n \bar{v}_{{p}}^2 {\tau}_0^2}) = {1} / {[{8 \pi \ ({\tau}_0 \left(n k_B T / m \right)^{1 / 2})^2}]} = 1/(8 \pi \gamma^2)$ by substituting the derived $\xi \approx \sqrt{2} \, \bar{v}_p \tau_0$ above into Eq.~\eqref{frenkeleq2D}, as obtained by directly numerical comparison.

One can now re-write the formula for the viscosity in the following form
\begin{equation}
\tcbhighmath[fuzzy halo=1mm with blue!50!white,arc=2pt,
  boxrule=0pt,frame hidden]{\eta =  \frac{m}{4\pi \tau _0} \frac{g(r)_{\max }}{g(r)_{{\min }}}}\quad ,\label{lala}
\end{equation}
in terms of the mass of each particle $m$, the high-temperature timescale $\tau_0$, and the pure short-range structural information based on the pair correlation function $g(r)$. 

Equation~\eqref{lala} completes in a sense Eyring's expression, $\eta= A \exp \left({\Delta G} / {k_B T}\right)$ \cite{eyring1935activated}, by providing a clear definition of the pre-factor $A$ and the energy barrier $\Delta G$. Indeed, in 2D liquids, in view of our results
\begin{equation}
    A \equiv \frac{m}{4 \pi \tau_0},\qquad \Delta G=\Delta w= k_B T \ln\left[{g(r)_{\max }} / {g(r)_{{\min }}}\right].
\end{equation}
In fact, $A = {m}/ {(4\pi \tau _0})$ is just the viscosity at high temperatures, \textit{i.e.} approaching the gas-like state, and the potential barrier $\Delta G$ is directly defined from the short-range order properties of the liquid. 

\begin{figure}[h]
    \centering
    \includegraphics[width=\linewidth]{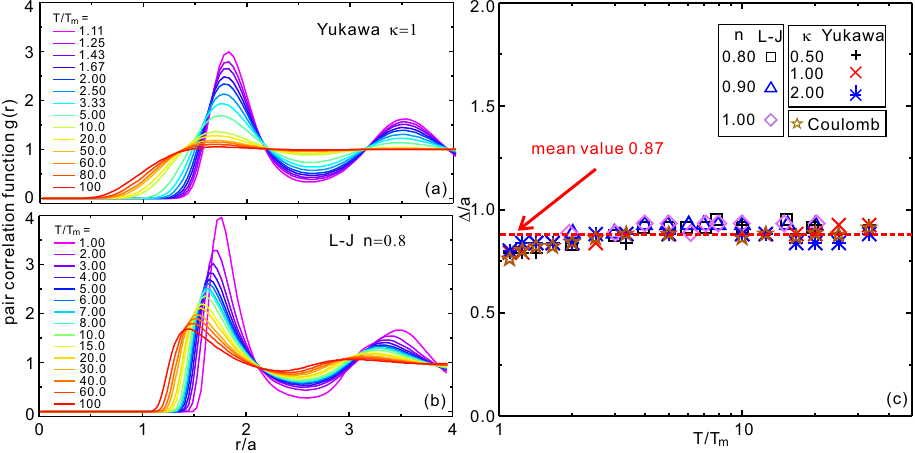}
    \caption{Calculated pair correlation functions $g(r)$ of 2D Yukawa \textbf{(a)} and L-J \textbf{(b)} fluids under various conditions, as well as the distance $\Delta$ between the first peak and first valley of $g(r)$ \textbf{(c)}.}
    \label{fig:5}
\end{figure}

After proving that the energy barrier governing liquid viscosity is the one between the first maximum and first minimum in $g(r)$, it comes naturally to identify the distance between these two as the length-scale $\xi$ associated to the potential hops in Frenkel's description of liquid dynamics. To confirm this hypothesis, we have computed the pair correlation functions $g(r)$ for 2D Yukawa and L-J fluids in a wide range of conditions, as presented in Figs.~\ref{fig:5}(a) and \ref{fig:5}(b).

We have then computed the distance $\Delta$ between the first peak and first valley in $g(r)$ as a function of the reduced temperature $T/T_m$, as presented in Fig.~\ref{fig:5}(c). Interestingly, when normalizing $\Delta$ using the Wigner-Seitz radius $a$, we find that this length-scale is approximately constant in temperature, showing mild deviations only at low temperatures. For both systems, we find that $\Delta \approx 0.87 a$. Using the values of $\delta$ from Eq.~\eqref{def}, we find that $\xi \approx 0.79 \Delta$ for the 2D Yukawa and OCP systems, and $\xi \approx 0.76 \Delta$ for the 2D L-J fluid. These results support the notion that the length scale $\xi$ in Frenkel's framework is temperature-independent and determined solely by the density $n$, which remains constant in Fig.~\ref{fig:5}. Moreover, they confirm that $\xi$ can, to a good approximation, be identified with $\Delta$ as defined from structure. This reinforces the idea that $\xi$ is effectively set by the average interparticle distance, consistent with the trends shown in Fig.~\ref{fig:5}(c) and with the assumption made in deriving Eq.~\eqref{maineq}.

Finally, it is worth noting that in the high-temperature regime, Eyring proposed a simple expression for viscosity, $\eta = n h$ \cite{10.1063/1.1749836}, where $h$ is Planck's constant and $n$ the particle number density. By invoking the uncertainty principle, $\Delta x \Delta p = h$, and identifying the relevant length scale with the microscopic length-scale $l_p$ and the momentum with the average particle momentum, this expression reduces to our prediction in the high-temperature limit where $\tau_{LC} = \tau_0$.

Moreover, it is interesting to compare the high-temperature limit of our viscosity formula $\eta=A$ with the expression from ideal kinetic theory, $\eta_{\text{kin}}^{\text{2D}} = \frac{1}{2} n m \langle v \rangle \, l_{\text{mfp}}$, where $\langle v \rangle$ is the average particle speed and $l_{\text{mfp}}$ is the mean free path. The average speed is given by $\langle v \rangle^2 = \pi k_B T / (2m)$, derived by averaging over the two-dimensional Maxwell–Boltzmann distribution. To first approximation, in 2D systems, the mean free path is given by $l_{\text{mfp}} = 1 / (2\sqrt{2} \, n a)$, where $n$ is the 2D areal number density of particles and $a$ is the Wigner–Seitz radius. Putting these together, the kinetic theory expression for viscosity in two dimensions becomes:
\begin{equation}
    \eta_{\text{kin}}^{\text{2D}} = \frac{\sqrt{\frac{\pi }{2}} m \bar{v}_p}{8 a} \approx 0.63\, \frac{m \bar{v}_p}{4a}.
\end{equation}
On the other hand, using that $\gamma = 0.18$ for the 2D Lennard-Jones system, and $\gamma = 0.20$ for the 2D Yukawa and OCP systems, our analysis yields:
\begin{align}
    & A \approx 0.72\, \frac{m \bar{v}_p}{4a} \qquad \text{for the 2D Lennard-Jones system}, \\
    & A \approx 0.65\, \frac{m \bar{v}_p}{4a} \qquad \text{for the 2D Yukawa and OCP systems}.
\end{align}

Thus, $A$ shares the same functional form as the ideal kinetic theory expression for viscosity, and the numerical prefactors agree within $ 3\%$ for the Yukawa and OCP systems, and within $12\%$ for the 2D Lennard-Jones system. This level of agreement is expected to be better for systems without attractive interactions, such as Yukawa and OCP models. It is worth noting that the kinetic theory formula used here is the ideal version, where collision integrals are taken as unity, as in the case of perfectly rigid particles with no interactions. It is possible that incorporating approximate collision integrals within the Chapman-Enskog formalism could further improve the match between our expression and kinetic theory, extending the latter slightly beyond the ideal limit.

We note that the same surprising agreement with the expression from kinetic theory was already remarked upon in Frenkel's book (page 200 in \cite{Frenkel1946}), where it is stated: ``\textit{It is interesting to note, however, that the magnitude of the viscosity coefficient which follows from our theory in the case of high temperatures lies very close to that which is obtained in the usual way from the kinetic theory of gases}.'' Our results in 2D are fully consistent with this observation.

Despite the interesting agreement with kinetic theory that emerges in the high-temperature limit, where $\tau_{LC} = \tau_0$, we emphasize that the physical principles underlying our formula are fundamentally different from those of kinetic theory. In particular, particle dynamics are not assumed to be collisional, $\xi$ does not represent a mean free path, and $\tau_0$ is not a collision time.

\section*{Discussion} 
In this work, we have considered the long-standing problem of deriving a microscopic and predictive formula for the shear viscosity of 2D simple fluids, a notorious challenge that can be summarized by the famous Landau argument that is ``\textit{impossible to derive any
general formulae giving a quantitative description of the
properties of a liquid}'' \cite{landau1980statistical}. Our findings prove that, in the dense liquid-like regime, an approximate microscopic formula for the viscosity can be found, in excellent agreement with the simulation data in several systems characterized by profoundly different particle interactions. In fact, our microscopic formula for the viscosity has been achieved not only at the particle level motion, but also from direct information of the short-range structural correlation of the liquid, which is encoded in its pair correlation function. Our derived viscosity equation connects the microscopic motion of particles, the macroscopic dynamics of collective shear waves, and the liquid structure of $g(r)$.

Our formula does not contradict Landau’s well-known argument, as it is not universally valid and breaks down in the dilute, gas-like regime. Nevertheless, it offers a useful and complementary framework to established theories such as Enskog theory and the hard-sphere fluid formalism, which are more appropriate for describing dilute systems. Importantly, our results complete three of the most successful frameworks to describe liquid dynamics and viscosity: Frenkel's theory, Maxwell's theory, and Eyring's theory. First, we propose and confirm the idea of identifying the microscopic Frenkel's time with the lifetime of local connectivity. Second, we find that the elastic length-scale below which collective shear waves propagate in liquids according to Maxwell model and $k$-gap theory can be directly connected to a single particle length-scale governing the diffusive transport of particle's momentum. Finally, we provide a precise definition of the undetermined parameters in Eyring's formula for viscosity and in particular we propose a simple method to obtain the hopping potential barrier from the short-range properties of the pair correlation function.

Do similar simple arguments apply to 3D liquids? Do our formulae provide an accurate estimate of the shear viscosity also for complex liquids and glass forming systems? This remains to be seen.

\section*{Methods}
\subsection*{Simulation Method for 2D simple liquid-like fluids}

We perform equilibrium molecular dynamics (MD) simulations of 2D Lennard-Jones (L-J), Yukawa, and Coulomb one-component plasma (OCP) liquid-like fluids. For all these three simple liquid-like fluids, the equation of motion for each particle is
\begin{equation}
\begin{aligned}
m\ddot{\mathbf{r}}_{i} = - \nabla \Sigma \phi_{i j},
\end{aligned}
\end{equation}
where $-\nabla \Sigma \phi_{i j}$ is the particle-particle interaction, while $\mathbf{r}_{i}$ is the position vector for the $i-$th particle. In our current , we always simulate $N = 4096$ particles constrained in a 2D simulation box with the length ratio of $L_x : L_y = 2 : \sqrt{3}$ with periodic boundary conditions. 

For each simulation run, first we integrate the equation of motion for all particles with a thermostat for $N_1$ steps, so that the simulation system reaches the specified conditions. Then, we turn off the thermostat to integrate the equation of motion for the next $N_2$ steps, and the obtained data are used for the data analysis presented in the main text. In our simulations, we specify the reduced temperature value $T/T_m$, where $T$ is the temperature of the simulated 2D system, while $T_m$ is the corresponding melting point. Also, we truncate the interparticle potential at $r_c$ to ensure that the potential energy of the simulation system does not change significantly with the increase of $r_c$ any more. We also verify that, for each simulation run, our time step is always chosen to be small enough, so that energy conservation is adequately obeyed.

\subsection*{2D L-J fluids}

For 2D L-J liquids, the interparticle interaction is
\begin{equation}
\begin{aligned}
\phi (r) = 4 \epsilon\left[(\sigma / r)^{12}-(\sigma / r)^6 \right],
\end{aligned}
\end{equation}
where $\epsilon$ and $\sigma$ are the energy and distance parameters. Here, we normalize the length and time using $\sigma$ and $\sqrt{m \sigma^2 / \epsilon}$, respectively. In our simulations of 2D L-J liquids, we specify the values of both the number density $n = N / A'$ and the reduced temperature $T / T_m$, where $A'$ is the area of the simulated box. 

Here are other simulation details. In our simulations, the number density are specified as $n = 0.8, 0.85, 0.9, 0.95,$ and 1.00. For each specified value of $n$, we vary the reduced temperature $T / T_m$ from 2 to 20. Note, we choose the melting points of 2D L-J systems $T_m$ for different $n$ values from~\cite{Massimo:2020}. In our 2D L-J simulations, the cutoff radius is chosen as $r_{c} = 2.5 \sigma$, while the corresponding integration steps are specified as $N_1 = 2 \times 10^{6}$ and $N_2 = 10^{9}$, respectively. 

\subsection*{2D Yukawa fluids}

For 2D Yukawa liquid-like fluids, the interparticle interaction is the Yukawa repulsion
\begin{equation}
\begin{aligned}
 \phi({r})=Q^{2} \exp \left(-r / \lambda_{D}\right) / 4 \pi \epsilon_{0} r,
\end{aligned}
\end{equation}
where $\lambda_{D}$ is the Debye length and $Q$ is the charge on each particle. Besides the reduced temperature $T / T_m$, we also use the screening parameter $\kappa = a / \lambda_D$ to characterize the simulated 2D Yukawa liquid-like fluids.

In order to mimic the conditions of most 2D dusty plasma experiments, we vary the $\kappa$ value from 0.75 to 2. For each $\kappa$ value, we vary the value of the reduced temperature $T / T_m$ from 1.11 to 10, where the values for the melting point $T_{m}$ are taken from~\cite{Hartmann:2005}. In our 2D Yukawa simulations, the cutoff radius is chosen as $r_{c} = 22a$, less than one half of each side of the simulation box. The integration steps are specified as $N_1 = 1 \times 10^{7}$ and $N_2 = 10^{8}$, respectively.

\subsection*{2D Coulomb OCP fluids}

For 2D Coulomb OCP liquid-like fluids, the interparticle interaction is 
\begin{equation}
\phi (r)=Q^{2} / 4 \pi \epsilon_{0} r.
\end{equation}
Unlike 2D L-J and Yukawa liquid-like fluids above, the interaction between particles in 2D Coulomb OCP liquid-like fluids is long-range. To avoid the Ewald summation~\cite{LeBard:2012}, we use the approximate potential~\cite{Fennell:2006}
\begin{equation}
\begin{aligned}
\phi (r)=\frac{Q^2}{4\pi \varepsilon _0}  {\left[\frac{\operatorname{erfc}(\alpha_1 r)}{r}-\frac{\operatorname{erfc}\left(\alpha_1 r_c\right)}{r_c}+\right.} \left.\left(\frac{\operatorname{erfc}\left(\alpha_1 r_c\right)}{r_c^2}+\frac{2 \alpha_1}{\sqrt{\pi}} \frac{\exp \left(-\alpha_1 ^2 r_c^2\right)}{r_c}\right)\left(r-r_c\right)\right],
\end{aligned}
\end{equation}
where $\alpha_1$ is the ``damping'' parameter, $r_c$ is the cutoff radius, and $\mathrm{erfc()}$ is the complementary error function. From previous studies, the choice of $\alpha_1 = 0.2$ enables the energy and forces of the simulated system to quickly converge to the long-range Coulomb system. In our 2D Coulomb OCP simulations, we choose $\alpha_1 = 0.2$ and $r_c = 10~a$, respectively. Other simulation details are the same as those for 2D Yukawa simulations described above. 

\subsection*{Lifetime of local connectivity from simulations}

In our current , to calculate the lifetime of local connectivity $\tau_{LC}$ of our simulated 2D liquid-like fluids, we need to track the neighbor list of all particles at each moment. For each studied particle $i$, its neighbors are defined as its pairing particle $j$ with their distance $r_{ij}$ less than the separation of the first minimum of the radial distribution function $g(r)$~\cite{PhysRevLett.110.205504,Ashwin:2015}. For example, in the initial configuration, there are $N(t_{0})$ neighbors for the studied particle $i$. As the time goes from the initial time $t_{0}$ to $t_{0}+t$, the neighbors of the studied particle $i$ change, i.e., some of the initial neighbors are not its neighbors any more. We may use $N(t_{0}+t)$ to label the number of the initial neighbors which are still its neighbors at the time of $t_{0}+t$. Thus, the lifetime of local connectivity $\tau_{L C}$ is defined as the time duration, relative to $t_{0}$, for the number of initial neighbors falls by $1$ in the ensemble average, \textit{i.e.}, $\left\langle N(t_{0})\right\rangle-\left\langle N(t_{0}+t)\right\rangle = 1$~\cite{PhysRevLett.110.205504,Ashwin:2015}, for all studied particles and varying the different initial times of $t_{0}$. In fact, if one neighbor leaves the studied particle $i$ for a while, then comes back as a neighbor again, it is still regarded as a new neighbor for the studied particle $i$. In summary, $\tau_{LC}$ can be regarded as the averaged time for the first of the initial neighbors of one particle $i$ goes beyond the distance of the first minimum of $g(r)$, i.e., the coordination number falls by $1$~\cite{Ashwin:2015}, or equivalently a new particle enter the range of one particle $i$ within the distance of the first minimum of $g(r)$ while none of the initial neighbors goes beyond. An extended analysis of the atomic connectivity network dynamics can be found in the Supplementary Material.

\section*{Data availability}
The datasets generated and analyzed during the current study are available upon reasonable request by contacting the corresponding authors. 

\section*{Code availability}
The codes that support the findings of this study is available upon reasonable request by contacting the corresponding authors. 

\section*{Acknowledgements}
This work was supported by the National Natural Science Foundation of China under Grants No. 12175159, No. 12305220, No. 12347110, the Excellent Postdoctoral Program of Jiangsu Province, the 1000 Youth Talents Plan, and the Priority Academic Program Development (PAPD) of Jiangsu Higher Education Institutions. M.B. acknowledges the support of the Shanghai Municipal Science and Technology Major Project (Grant No.2019SHZDZX01) and the sponsorship from the Yangyang Development Fund.

\section*{Author contributions}
Y. F. and D. H. conceived the idea of this project. Y. F. and M. B. supervised the project. D. H., S. L. and C. L. performed the simulations and the data analysis. All authors contributed to the writing of the manuscript and the theoretical interpretation of the results.
\section*{Competing interests}
The authors declare no competing interests.

\section*{Supplementary Material}
\subsection*{Specifications of the regime under investigation}
In this section, we provide a more detailed overview of the parameter space explored in the 2D fluid systems analyzed in our study.
\begin{figure}[h]
    \centering
    \includegraphics{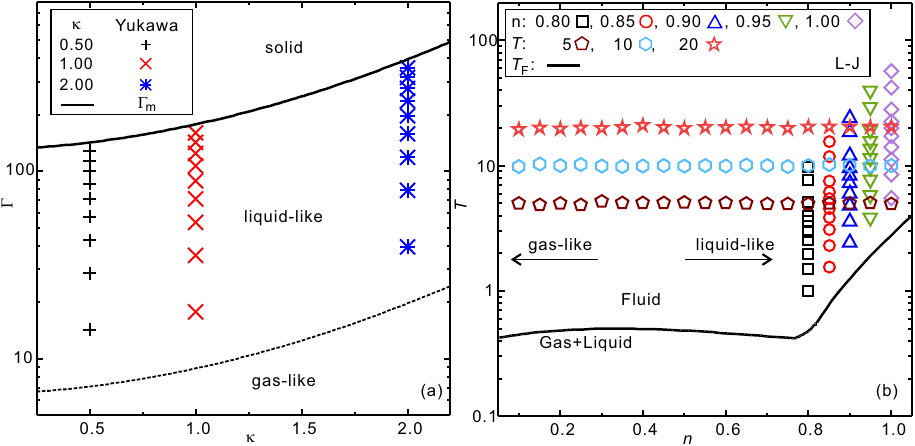}
    \caption{Regimes under  within the phase diagram of 2D Yukawa \textbf{(a)} and 2D L-J \textbf{(b)} systems. The horizontal scans in the L-J phase diagram (b) correspond to the data in Fig.~\ref{fig:rev}, while the vertical ones refer to all the other analyses performed in the main text. }
    \label{fig:r4}
\end{figure}

\subsubsection*{2D Yukawa and OCP systems}
For 2D Yukawa and OCP systems, as for any model with purely repulsive interactions, a first-order thermodynamic separation between the liquid and gas phases does not exist. Nevertheless, similar to supercritical fluids, a dynamical crossover between a liquid-like and a gas-like phase can be defined and probed with different physical observables, see for example \cite{PhysRevResearch.5.013149}.

The phase diagram of the 2D Yukawa system is shown in Fig.~\ref{fig:r4}(a) as a function of the parameter $\Gamma$ and $\kappa$ that play respectively the role of inverse temperature and inverse density. The colored symbols represented the regions analyzed and discussed in the main text. As evident from the figure, all the data points lie within the liquid-like fluid phase and below the melting temperature $T_m$ (solid line in Fig.~\ref{fig:r4}(a)).

For Yukawa systems, $T_m$ is adopted from Ref.\cite{Hartmann:2005} and determined using the melting criterion proposed in Ref.\cite{PhysRevLett.82.5293}, which is based on the bond-angle correlation factor. Finally, for the OCP system, $T_m$ characterizes the phase boundary for the electron-liquid to electron- crystal phase transition, and it is taken from \cite{PhysRevLett.42.795} (see also References therein for other theoretical estimates).

\subsubsection*{2D L-J system}
The phase diagram of the 2D L-J system is shown in Fig.~\ref{fig:r4}(b), plotted in the temperature–density plane. For this system, the melting temperature $T_m$ (solid line) corresponds to the boundary between the fluid and coexistence regions, as reported in Ref.\cite{Massimo:2020}. The vertical scans represent the data analyzed in the main text, while the horizontal scans correspond to the density-dependent data shown in Fig.~\ref{fig:rev} in the main text. As clearly seen, all datasets lie within the fluid phase. Specifically, the data discussed in the main text fall entirely within the dense, liquid-like regime, whereas the horizontal scans extend across both gas-like and liquid-like regimes.

\subsection*{Atomic connectivity network dynamics}
Our discussion of viscosity in the main text is grounded in the properties of the atomic connectivity network. In this section, we present a more detailed analysis of its dynamics. First, we compute the autocorrelation function 
\begin{equation}
    N_c(t)\equiv \langle n(t)n(0)\rangle,\label{auto}
\end{equation}
where $n$ denotes the number of initial neighbors 
that are preserved during a time interval $t$.

\begin{figure}[ht]
    \centering
    \includegraphics{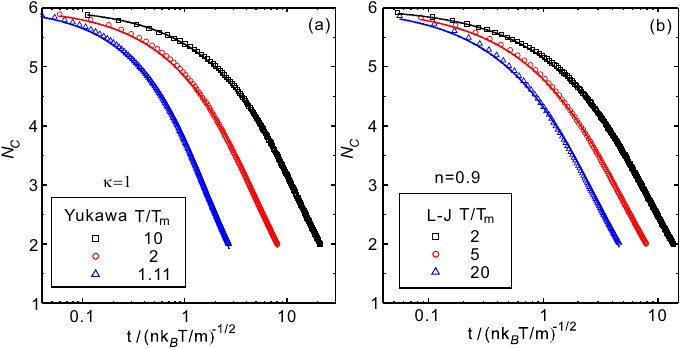}
    \caption{Number of neighbors autocorrelation function $N_c(t)$, Eq.~\eqref{auto}, as a function of dimensionless time for 2D Yukawa \textbf{(a)} and 2D L-J \textbf{(b)} systems. Different colors correspond to different temperature conditions. Solid lines represent the fits to Eq.~\eqref{st}.}
    \label{fig:r1}
\end{figure}

\begin{figure}[h]
    \centering
    \includegraphics[width=\linewidth]{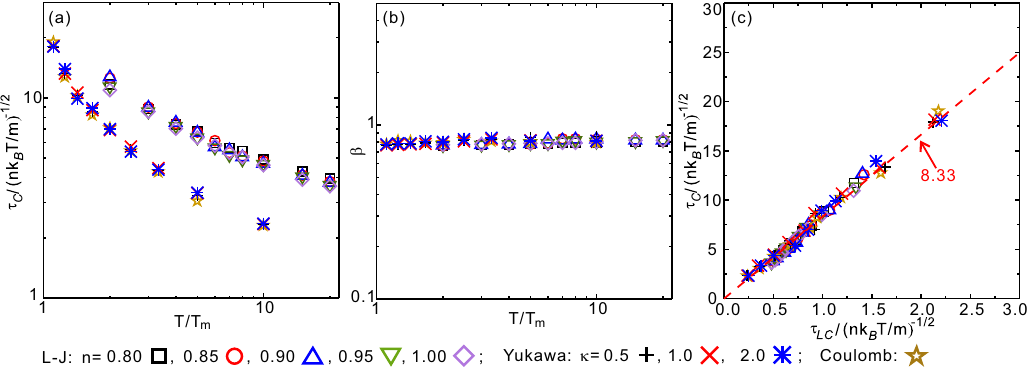}
    \caption{\textbf{(a)} Relaxation time $\tau_c$ as a function of temperature obtained by fitting with a stretched exponential function, Eq.~\eqref{st}, to the autocorrelation function $N_c(t)$. \textbf{(b)} The stretched exponential parameter $\beta$ obtained from the same procedure. \textbf{(c)} Demonstration of the linear proportionality between the relaxation time $\tau_c$ and the lifetime of local connectivity $\tau_{LC}$.}
    \label{fig:r2}
\end{figure}

In Fig.~\ref{fig:r1}, we present the autocorrelation function $N_c(t)$ for 2D Yukawa and 2D L-J systems under different conditions. Our results indicate that $N_c$ decays in time from its initial value $N_c(t=0)=6$. Most importantly, in all cases, we observe that the decay becomes faster as the temperature increases. To quantify this behavior in more detail, we fit the data using a stretched exponential function:
\begin{equation}
    N_c(t)=N_0 \exp \left(-(t/\tau_c)^\beta\right),\label{st}
\end{equation}
which provides an excellent description of our simulation results. 

The results of the fit are shown in Fig.~\ref{fig:r2}. As anticipated, the relaxation time $\tau_c$ decreases with increasing temperature, independently of the interparticle potential. This indicates that the microscopic process of losing a neighbor becomes slower and rarer as the temperature decreases. We also observe that the decay is not perfectly exponential (see panel (b) of Fig.~\ref{fig:r2}), as evidenced by a stretching exponent $\beta$ slightly smaller than 1. Interestingly, $\beta$ appears, to first approximation, to be independent of both the thermodynamic conditions and the interaction potential. This surprising universality merits further s in the future.

Finally, we highlight another intriguing observation related to the relaxation timescale $\tau_c$, which is extracted from the autocorrelation function $N_c(t)$. As shown in Fig.~\ref{fig:r2}(c) for both 2D L-J and Yukawa systems, this timescale $\tau_c$ is directly proportional to the lifetime of local connectivity, $\tau_{LC}$, that appears in our viscosity formula. 

\subsection*{Long time tails and system size dependence}
\begin{figure}[ht]
    \centering
    \includegraphics[width=0.9\linewidth]{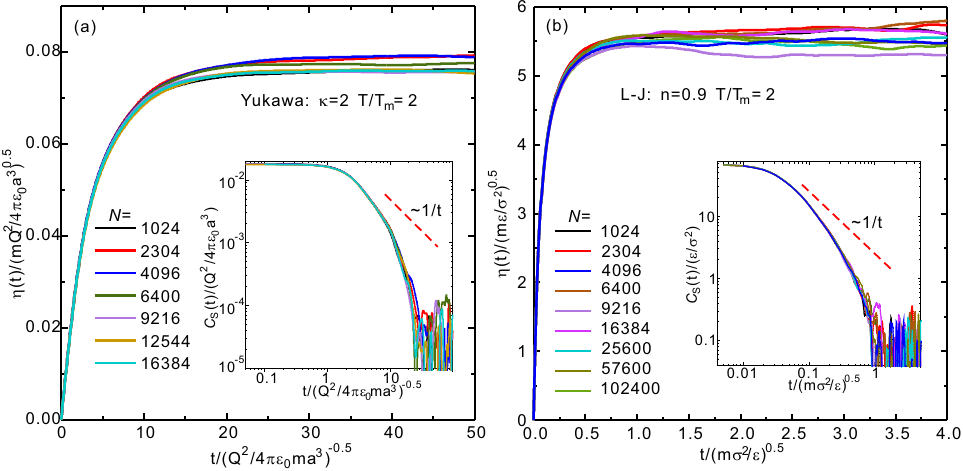}
    \caption{Time dependent viscosity $\eta(t)=\int_0^t C_S(t')dt'$, with $C_s(t)$ the shear stress autocorrelation function, for different system sizes $N$ in both 2D Yukawa \textbf{(a)} and 2D L-J \textbf{(b)} systems. The insets prove the absence of the $1/t$ decay characteristic of long-time tails.}
    \label{fig:notail}
\end{figure}

\begin{figure}[ht]
    \centering
    \includegraphics{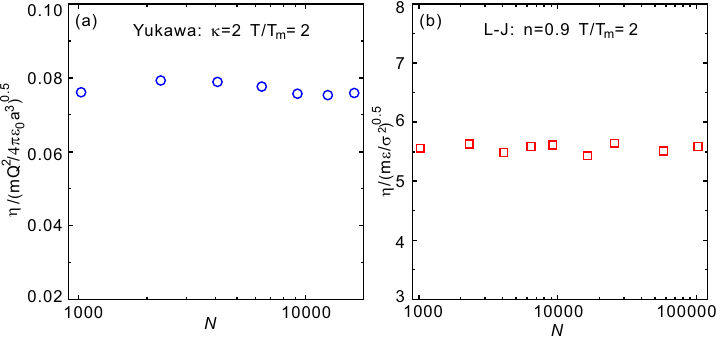}
    \caption{Size dependence of the viscosity $\eta$ for 2D Yukawa \textbf{(a)} and 2D L-J \textbf{(b)} systems. Within statistical error, no clear $N$ dependence is observed, consistent with the absence of long time tails demonstrated in Fig.~\ref{fig:notail}.}
    \label{fig:const}
\end{figure}

As first reported in the seminal work of Alder and Wainwright on hard-sphere fluids \cite{PhysRevA.1.18} (see also \cite{PhysRevLett.25.1254}), autocorrelation functions in 2D systems can exhibit a characteristic $1/t$ decay, known as the hydrodynamic long-time tails. This phenomenon leads to divergences in transport coefficients, such as the diffusion constant, viscosity, and thermal conductivity, when computed using the Green-Kubo formalism. As a result, these coefficients become dependent on system size and are no longer intrinsic properties of the system.

While long-time tails are well established in hard-sphere fluids, their universality, particularly in systems with soft interactions, like those considered in this work, remains an open question. In fact, earlier studies \cite{PhysRevE.52.6123,PhysRevLett.93.155004} found no evidence of long-time tails for viscosity at least in certain 2D soft matter systems, such as 2D Yukawa fluids.

In Fig.~\ref{fig:notail}, we present the shear stress autocorrelation function $C_S(t)$ for different system sizes, ranging from $N=1024$ to 
$N=102400$, for both 2D L-J and 2D Yukawa systems. As illustrated in the insets, we observe no sign of a $1/t$ decay, which would indicate the presence of long-time tails. For completeness, we also show the time-dependent viscosity, $\eta(t)=\int_0^t C_S(t')dt'$, from which the final viscosity $\eta$ is extracted. In Fig.~\ref{fig:const}, we display the resulting viscosity as a function of system size $N$ for both 2D L-J and 2D Yukawa systems. Within statistical error, we find no significant system size dependence, and the viscosity remains finite and well-defined.

Although a more extensive analysis is needed for a definitive conclusion, our results suggest that, at least in the systems studied here, no evidence of the hydrodynamic long-time tail for viscosity is present. We therefore conclude that viscosity remains a well-defined transport coefficient in these 2D soft-interaction systems.


\begin{thebibliography}{10}
\urlstyle{rm}
\expandafter\ifx\csname url\endcsname\relax
  \def\url#1{\texttt{#1}}\fi
\expandafter\ifx\csname urlprefix\endcsname\relax\def\urlprefix{URL }\fi
\expandafter\ifx\csname doiprefix\endcsname\relax\def\doiprefix{DOI: }\fi
\providecommand{\bibinfo}[2]{#2}
\providecommand{\eprint}[2][]{\url{#2}}

\bibitem{landau1987fluid}
\bibinfo{author}{Landau, L.~D.} \& \bibinfo{author}{Lifshitz, E.~M.}
\newblock \emph{\bibinfo{title}{Fluid Mechanics (Second Edition)}} (\bibinfo{publisher}{Pergamon}, \bibinfo{year}{1987}).

\bibitem{hansen2013theory}
\bibinfo{author}{Hansen, J.-P.} \& \bibinfo{author}{McDonald, I.~R.}
\newblock \emph{\bibinfo{title}{Theory of simple liquids: with applications to soft matter}} (\bibinfo{publisher}{Academic press}, \bibinfo{year}{2013}).

\bibitem{Evans_Morriss_2008}
\bibinfo{author}{Evans, D.~J.} \& \bibinfo{author}{Morriss, G.}
\newblock \emph{\bibinfo{title}{Statistical Mechanics of Nonequilibrium Liquids}} (\bibinfo{publisher}{Cambridge University Press}, \bibinfo{year}{2008}), \bibinfo{edition}{2} edn.

\bibitem{Evans01071978}
\bibinfo{author}{Evans, D.~J.} \& \bibinfo{author}{and, W. B.~S.}
\newblock \bibinfo{journal}{\bibinfo{title}{Transport properties of homonuclear diatomics}}.
\newblock {\emph{\JournalTitle{Molecular Physics}}} \textbf{\bibinfo{volume}{36}}, \bibinfo{pages}{161--176}, \doiprefix\url{10.1080/00268977800101491} (\bibinfo{year}{1978}).

\bibitem{loeb2004kinetic}
\bibinfo{author}{Loeb, L.~B.}
\newblock \emph{\bibinfo{title}{The kinetic theory of gases}} (\bibinfo{publisher}{Courier Corporation}, \bibinfo{year}{2004}).

\bibitem{Frenkel1946}
\bibinfo{author}{Frenkel, J.}
\newblock \emph{\bibinfo{title}{Kinetic theory of liquids}}.
\newblock International series of monographs on physics (\bibinfo{publisher}{Clarendon Press Oxford}, \bibinfo{address}{Oxford}, \bibinfo{year}{1946}).

\bibitem{eyring1935activated}
\bibinfo{author}{Eyring, H.}
\newblock \bibinfo{journal}{\bibinfo{title}{The activated complex in chemical reactions}}.
\newblock {\emph{\JournalTitle{The Journal of Chemical Physics}}} \textbf{\bibinfo{volume}{3}}, \bibinfo{pages}{107--115} (\bibinfo{year}{1935}).

\bibitem{osti_5437529}
\bibinfo{author}{Touloukian, Y.~S.}, \bibinfo{author}{Saxena, S.~C.} \& \bibinfo{author}{Hestermans, P.}
\newblock \emph{\bibinfo{title}{Thermophysical properties of matter - the TPRC data series. Volume 11. Viscosity. (Reannouncement). Data book}} (\bibinfo{year}{1975}).

\bibitem{PhysRevLett.110.205504}
\bibinfo{author}{Iwashita, T.}, \bibinfo{author}{Nicholson, D.~M.} \& \bibinfo{author}{Egami, T.}
\newblock \bibinfo{journal}{\bibinfo{title}{Elementary excitations and crossover phenomenon in liquids}}.
\newblock {\emph{\JournalTitle{Phys. Rev. Lett.}}} \textbf{\bibinfo{volume}{110}}, \bibinfo{pages}{205504}, \doiprefix\url{10.1103/PhysRevLett.110.205504} (\bibinfo{year}{2013}).

\bibitem{PhysRevE.98.022604}
\bibinfo{author}{Shinohara, Y.} \emph{et~al.}
\newblock \bibinfo{journal}{\bibinfo{title}{Viscosity and real-space molecular motion of water: Observation with inelastic x-ray scattering}}.
\newblock {\emph{\JournalTitle{Phys. Rev. E}}} \textbf{\bibinfo{volume}{98}}, \bibinfo{pages}{022604}, \doiprefix\url{10.1103/PhysRevE.98.022604} (\bibinfo{year}{2018}).

\bibitem{10.1063/1.4789306}
\bibinfo{author}{Levashov, V.~A.}, \bibinfo{author}{Morris, J.~R.} \& \bibinfo{author}{Egami, T.}
\newblock \bibinfo{journal}{\bibinfo{title}{The origin of viscosity as seen through atomic level stress correlation function}}.
\newblock {\emph{\JournalTitle{The Journal of Chemical Physics}}} \textbf{\bibinfo{volume}{138}}, \bibinfo{pages}{044507}, \doiprefix\url{10.1063/1.4789306} (\bibinfo{year}{2013}).

\bibitem{PhysRevE.98.063005}
\bibinfo{author}{Bellissard, J.} \& \bibinfo{author}{Egami, T.}
\newblock \bibinfo{journal}{\bibinfo{title}{Simple theory of viscosity in liquids}}.
\newblock {\emph{\JournalTitle{Phys. Rev. E}}} \textbf{\bibinfo{volume}{98}}, \bibinfo{pages}{063005}, \doiprefix\url{10.1103/PhysRevE.98.063005} (\bibinfo{year}{2018}).

\bibitem{Trachenko_2016}
\bibinfo{author}{Trachenko, K.} \& \bibinfo{author}{Brazhkin, V.~V.}
\newblock \bibinfo{journal}{\bibinfo{title}{Collective modes and thermodynamics of the liquid state}}.
\newblock {\emph{\JournalTitle{Reports on Progress in Physics}}} \textbf{\bibinfo{volume}{79}}, \bibinfo{pages}{016502}, \doiprefix\url{10.1088/0034-4885/79/1/016502} (\bibinfo{year}{2015}).

\bibitem{BAGGIOLI20201}
\bibinfo{author}{Baggioli, M.}, \bibinfo{author}{Vasin, M.}, \bibinfo{author}{Brazhkin, V.} \& \bibinfo{author}{Trachenko, K.}
\newblock \bibinfo{journal}{\bibinfo{title}{Gapped momentum states}}.
\newblock {\emph{\JournalTitle{Physics Reports}}} \textbf{\bibinfo{volume}{865}}, \bibinfo{pages}{1--44}, \doiprefix\url{https://doi.org/10.1016/j.physrep.2020.04.002} (\bibinfo{year}{2020}).
\newblock \bibinfo{note}{Gapped momentum states}.

\bibitem{peluso2024viscosityliquidsdualmodel}
\bibinfo{author}{Peluso, F.}
\newblock \bibinfo{journal}{\bibinfo{title}{The viscosity of liquids in the dual model}}.
\newblock {\emph{\JournalTitle{Thermo}}} \textbf{\bibinfo{volume}{4}}, \bibinfo{pages}{508--539}, \doiprefix\url{10.3390/thermo4040028} (\bibinfo{year}{2024}).

\bibitem{doi:10.1098/rspa.1947.0088}
\bibinfo{author}{Born, M.} \& \bibinfo{author}{Green, H.~S.}
\newblock \bibinfo{journal}{\bibinfo{title}{A general kinetic theory of liquids iii. dynamical properties}}.
\newblock {\emph{\JournalTitle{Proceedings of the Royal Society of London. Series A. Mathematical and Physical Sciences}}} \textbf{\bibinfo{volume}{190}}, \bibinfo{pages}{455--474}, \doiprefix\url{10.1098/rspa.1947.0088} (\bibinfo{year}{1947}).

\bibitem{EVANS1980321}
\bibinfo{author}{Evans, D.~J.} \& \bibinfo{author}{Watts, R.}
\newblock \bibinfo{journal}{\bibinfo{title}{Shear-dependent viscosity in simple fluids}}.
\newblock {\emph{\JournalTitle{Chemical Physics}}} \textbf{\bibinfo{volume}{48}}, \bibinfo{pages}{321--327}, \doiprefix\url{https://doi.org/10.1016/0301-0104(80)80063-2} (\bibinfo{year}{1980}).

\bibitem{due}
\bibinfo{author}{Koo, H.-M.} \& \bibinfo{author}{Hess, S.}
\newblock \bibinfo{journal}{\bibinfo{title}{{Pair-correlation function of a fluid undergoing a simple shear flow: Solution of the Kirkwood-Smoluchowski equation}}}.
\newblock {\emph{\JournalTitle{Physica A: Statistical Mechanics and its Applications}}} \textbf{\bibinfo{volume}{145}}, \bibinfo{pages}{361--407}, \doiprefix\url{10.1016/0378-4371(87)90002-1} (\bibinfo{year}{1987}).

\bibitem{PhysRevE.108.044101}
\bibinfo{author}{Zaccone, A.}
\newblock \bibinfo{journal}{\bibinfo{title}{General theory of the viscosity of liquids and solids from nonaffine particle motions}}.
\newblock {\emph{\JournalTitle{Phys. Rev. E}}} \textbf{\bibinfo{volume}{108}}, \bibinfo{pages}{044101}, \doiprefix\url{10.1103/PhysRevE.108.044101} (\bibinfo{year}{2023}).

\bibitem{huang2024microscopicoriginliquidviscosity}
\bibinfo{author}{Huang, L.-Z.}, \bibinfo{author}{Cui, B.}, \bibinfo{author}{Vaibhav, V.}, \bibinfo{author}{Baggioli, M.} \& \bibinfo{author}{Wang, Y.-J.}
\newblock \bibinfo{title}{Microscopic origin of liquid viscosity from unstable localized modes} (\bibinfo{year}{2024}).
\newblock \eprint{2408.07937}.

\bibitem{doi:10.1126/sciadv.1603079}
\bibinfo{author}{Iwashita, T.} \emph{et~al.}
\newblock \bibinfo{journal}{\bibinfo{title}{Seeing real-space dynamics of liquid water through inelastic x-ray scattering}}.
\newblock {\emph{\JournalTitle{Science Advances}}} \textbf{\bibinfo{volume}{3}}, \bibinfo{pages}{e1603079}, \doiprefix\url{10.1126/sciadv.1603079} (\bibinfo{year}{2017}).

\bibitem{D0CP01560A}
\bibinfo{author}{Yahya, A.} \emph{et~al.}
\newblock \bibinfo{journal}{\bibinfo{title}{Molecular origins of bulk viscosity in liquid water}}.
\newblock {\emph{\JournalTitle{Phys. Chem. Chem. Phys.}}} \textbf{\bibinfo{volume}{22}}, \bibinfo{pages}{9494--9502}, \doiprefix\url{10.1039/D0CP01560A} (\bibinfo{year}{2020}).

\bibitem{PhysRevResearch.4.033064}
\bibinfo{author}{Huang, D.}, \bibinfo{author}{Lu, S.}, \bibinfo{author}{Murillo, M.~S.} \& \bibinfo{author}{Feng, Y.}
\newblock \bibinfo{journal}{\bibinfo{title}{Origin of viscosity at individual particle level in yukawa liquids}}.
\newblock {\emph{\JournalTitle{Phys. Rev. Res.}}} \textbf{\bibinfo{volume}{4}}, \bibinfo{pages}{033064}, \doiprefix\url{10.1103/PhysRevResearch.4.033064} (\bibinfo{year}{2022}).

\bibitem{yu2025understandingflowbehaviorssupercooled}
\bibinfo{author}{Yu, D.-X.}, \bibinfo{author}{Zeng, K.-Q.} \& \bibinfo{author}{Wang, Z.}
\newblock \bibinfo{title}{Understanding flow behaviors of supercooled liquids by embodying solid-liquid duality at particle level} (\bibinfo{year}{2025}).
\newblock \eprint{2506.03818}.

\bibitem{jiang2024experimental}
\bibinfo{author}{Jiang, C.}, \bibinfo{author}{Zheng, Z.}, \bibinfo{author}{Chen, Y.}, \bibinfo{author}{Baggioli, M.} \& \bibinfo{author}{Zhang, J.}
\newblock \bibinfo{journal}{\bibinfo{title}{Experimental observation of gapped shear waves and liquid-like to gas-like dynamical crossover in active granular matter}}.
\newblock {\emph{\JournalTitle{Communications Physics}}} \textbf{\bibinfo{volume}{8}}, \bibinfo{pages}{82}, \doiprefix\url{10.1038/s42005-025-02008-1} (\bibinfo{year}{2025}).

\bibitem{10.1063/1.4834696}
\bibinfo{author}{Balboa~Usabiaga, F.}, \bibinfo{author}{Xie, X.}, \bibinfo{author}{Delgado-Buscalioni, R.} \& \bibinfo{author}{Donev, A.}
\newblock \bibinfo{journal}{\bibinfo{title}{The stokes-einstein relation at moderate schmidt number}}.
\newblock {\emph{\JournalTitle{The Journal of Chemical Physics}}} \textbf{\bibinfo{volume}{139}}, \bibinfo{pages}{214113}, \doiprefix\url{10.1063/1.4834696} (\bibinfo{year}{2013}).

\bibitem{PhysRevA.1.18}
\bibinfo{author}{Alder, B.~J.} \& \bibinfo{author}{Wainwright, T.~E.}
\newblock \bibinfo{journal}{\bibinfo{title}{Decay of the velocity autocorrelation function}}.
\newblock {\emph{\JournalTitle{Phys. Rev. A}}} \textbf{\bibinfo{volume}{1}}, \bibinfo{pages}{18--21}, \doiprefix\url{10.1103/PhysRevA.1.18} (\bibinfo{year}{1970}).

\bibitem{PhysRevLett.25.1254}
\bibinfo{author}{Ernst, M.~H.}, \bibinfo{author}{Hauge, E.~H.} \& \bibinfo{author}{van Leeuwen, J. M.~J.}
\newblock \bibinfo{journal}{\bibinfo{title}{Asymptotic time behavior of correlation functions}}.
\newblock {\emph{\JournalTitle{Phys. Rev. Lett.}}} \textbf{\bibinfo{volume}{25}}, \bibinfo{pages}{1254--1256}, \doiprefix\url{10.1103/PhysRevLett.25.1254} (\bibinfo{year}{1970}).

\bibitem{PhysRevE.52.6123}
\bibinfo{author}{Gravina, D.}, \bibinfo{author}{Ciccotti, G.} \& \bibinfo{author}{Holian, B.~L.}
\newblock \bibinfo{journal}{\bibinfo{title}{Linear and nonlinear viscous flow in two-dimensional fluids}}.
\newblock {\emph{\JournalTitle{Phys. Rev. E}}} \textbf{\bibinfo{volume}{52}}, \bibinfo{pages}{6123--6128}, \doiprefix\url{10.1103/PhysRevE.52.6123} (\bibinfo{year}{1995}).

\bibitem{PhysRevLett.93.155004}
\bibinfo{author}{Nosenko, V.} \& \bibinfo{author}{Goree, J.}
\newblock \bibinfo{journal}{\bibinfo{title}{Shear flows and shear viscosity in a two-dimensional yukawa system (dusty plasma)}}.
\newblock {\emph{\JournalTitle{Phys. Rev. Lett.}}} \textbf{\bibinfo{volume}{93}}, \bibinfo{pages}{155004}, \doiprefix\url{10.1103/PhysRevLett.93.155004} (\bibinfo{year}{2004}).

\bibitem{1981PhyA..106..226B}
\bibinfo{author}{{Barker}, J.~A.}, \bibinfo{author}{{Henderson}, D.} \& \bibinfo{author}{{Abraham}, F.~F.}
\newblock \bibinfo{journal}{\bibinfo{title}{{Phase diagram of the two-dimensional Lennard-Jones system; Evidence for first-order transitions}}}.
\newblock {\emph{\JournalTitle{Physica A Statistical Mechanics and its Applications}}} \textbf{\bibinfo{volume}{106}}, \bibinfo{pages}{226--238}, \doiprefix\url{10.1016/0378-4371(81)90222-3} (\bibinfo{year}{1981}).

\bibitem{Melzer:1996}
\bibinfo{author}{Melzer, A.}, \bibinfo{author}{Homann, A.} \& \bibinfo{author}{Piel, A.}
\newblock \bibinfo{journal}{\bibinfo{title}{Experimental investigation of the melting transition of the plasma crystal}}.
\newblock {\emph{\JournalTitle{Phys. Rev. E}}} \textbf{\bibinfo{volume}{53}}, \bibinfo{pages}{2757--2766}, \doiprefix\url{10.1103/PhysRevE.53.2757} (\bibinfo{year}{1996}).

\bibitem{RevModPhys1353}
\bibinfo{author}{Morfill, G.~E.} \& \bibinfo{author}{Ivlev, A.~V.}
\newblock \bibinfo{journal}{\bibinfo{title}{Complex plasmas: An interdisciplinary research field}}.
\newblock {\emph{\JournalTitle{Rev. Mod. Phys.}}} \textbf{\bibinfo{volume}{81}}, \bibinfo{pages}{1353}, \doiprefix\url{10.1103/RevModPhys.81.1353} (\bibinfo{year}{2009}).

\bibitem{PhysRevLett145003}
\bibinfo{author}{Donk\'o, Z.}, \bibinfo{author}{Goree, J.}, \bibinfo{author}{Hartmann, P.} \& \bibinfo{author}{Kutasi, K.}
\newblock \bibinfo{journal}{\bibinfo{title}{Shear viscosity and shear thinning in two-dimensional yukawa liquids}}.
\newblock {\emph{\JournalTitle{Phys. Rev. Lett.}}} \textbf{\bibinfo{volume}{96}}, \bibinfo{pages}{145003}, \doiprefix\url{10.1103/PhysRevLett.96.145003} (\bibinfo{year}{2006}).

\bibitem{PhysRevLett065003}
\bibinfo{author}{Daligault, J.}
\newblock \bibinfo{journal}{\bibinfo{title}{Liquid-state properties of a one-component plasma}}.
\newblock {\emph{\JournalTitle{Phys. Rev. Lett.}}} \textbf{\bibinfo{volume}{96}}, \bibinfo{pages}{065003}, \doiprefix\url{10.1103/PhysRevLett.96.145003} (\bibinfo{year}{2006}).

\bibitem{PhysRevLett235001}
\bibinfo{author}{Baalrud, S.~D.} \& \bibinfo{author}{Daligault, J.}
\newblock \bibinfo{journal}{\bibinfo{title}{Effective potential theory for transport coefficients across coupling regimes}}.
\newblock {\emph{\JournalTitle{Phys. Rev. Lett.}}} \textbf{\bibinfo{volume}{110}}, \bibinfo{pages}{235001}, \doiprefix\url{10.1103/PhysRevLett.110.235001} (\bibinfo{year}{2013}).

\bibitem{Rosenfeld:2001}
\bibinfo{author}{Rosenfeld, Y.}
\newblock \bibinfo{journal}{\bibinfo{title}{Quasi-universal melting-temperature scaling of transport coefficients in yukawa systems}}.
\newblock {\emph{\JournalTitle{Journal of Physics: Condensed Matter}}} \textbf{\bibinfo{volume}{13}}, \bibinfo{pages}{L39}, \doiprefix\url{10.1088/0953-8984/13/2/101} (\bibinfo{year}{2001}).

\bibitem{Ashwin:2015}
\bibinfo{author}{Ashwin, J.} \& \bibinfo{author}{Sen, A.}
\newblock \bibinfo{journal}{\bibinfo{title}{Microscopic origin of shear relaxation in a model viscoelastic liquid}}.
\newblock {\emph{\JournalTitle{Phys. Rev. Lett.}}} \textbf{\bibinfo{volume}{114}}, \bibinfo{pages}{055002}, \doiprefix\url{10.1103/PhysRevLett.114.055002} (\bibinfo{year}{2015}).

\bibitem{Massimo:2020}
\bibinfo{author}{Li, Y.-W.} \& \bibinfo{author}{Ciamarra, M.~P.}
\newblock \bibinfo{journal}{\bibinfo{title}{Phase behavior of lennard-jones particles in two dimensions}}.
\newblock {\emph{\JournalTitle{Phys. Rev. E}}} \textbf{\bibinfo{volume}{102}}, \bibinfo{pages}{062101}, \doiprefix\url{10.1103/PhysRevE.102.062101} (\bibinfo{year}{2020}).

\bibitem{maxwell1867iv}
\bibinfo{author}{Maxwell, J.~C.}
\newblock \bibinfo{journal}{\bibinfo{title}{{IV. On the dynamical theory of gases}}}.
\newblock {\emph{\JournalTitle{Philosophical transactions of the Royal Society of London}}} \bibinfo{pages}{49--88} (\bibinfo{year}{1867}).

\bibitem{PhysRevE.105.024602}
\bibinfo{author}{Baggioli, M.}, \bibinfo{author}{Landry, M.} \& \bibinfo{author}{Zaccone, A.}
\newblock \bibinfo{journal}{\bibinfo{title}{Deformations, relaxation, and broken symmetries in liquids, solids, and glasses: A unified topological field theory}}.
\newblock {\emph{\JournalTitle{Phys. Rev. E}}} \textbf{\bibinfo{volume}{105}}, \bibinfo{pages}{024602}, \doiprefix\url{10.1103/PhysRevE.105.024602} (\bibinfo{year}{2022}).

\bibitem{PhysRevLett.118.215502}
\bibinfo{author}{Yang, C.}, \bibinfo{author}{Dove, M.~T.}, \bibinfo{author}{Brazhkin, V.~V.} \& \bibinfo{author}{Trachenko, K.}
\newblock \bibinfo{journal}{\bibinfo{title}{Emergence and evolution of the $k$ gap in spectra of liquid and supercritical states}}.
\newblock {\emph{\JournalTitle{Phys. Rev. Lett.}}} \textbf{\bibinfo{volume}{118}}, \bibinfo{pages}{215502}, \doiprefix\url{10.1103/PhysRevLett.118.215502} (\bibinfo{year}{2017}).

\bibitem{PhysRevLett.97.115001}
\bibinfo{author}{Nosenko, V.}, \bibinfo{author}{Goree, J.} \& \bibinfo{author}{Piel, A.}
\newblock \bibinfo{journal}{\bibinfo{title}{Cutoff wave number for shear waves in a two-dimensional yukawa system (dusty plasma)}}.
\newblock {\emph{\JournalTitle{Phys. Rev. Lett.}}} \textbf{\bibinfo{volume}{97}}, \bibinfo{pages}{115001}, \doiprefix\url{10.1103/PhysRevLett.97.115001} (\bibinfo{year}{2006}).

\bibitem{PhysRevLett.92.065001}
\bibinfo{author}{Kalman, G.~J.}, \bibinfo{author}{Hartmann, P.}, \bibinfo{author}{Donk\'o, Z.} \& \bibinfo{author}{Rosenberg, M.}
\newblock \bibinfo{journal}{\bibinfo{title}{Two-dimensional yukawa liquids: Correlation and dynamics}}.
\newblock {\emph{\JournalTitle{Phys. Rev. Lett.}}} \textbf{\bibinfo{volume}{92}}, \bibinfo{pages}{065001}, \doiprefix\url{10.1103/PhysRevLett.92.065001} (\bibinfo{year}{2004}).

\bibitem{10.1063/1.1749836}
\bibinfo{author}{Eyring, H.}
\newblock \bibinfo{journal}{\bibinfo{title}{Viscosity, plasticity, and diffusion as examples of absolute reaction rates}}.
\newblock {\emph{\JournalTitle{The Journal of Chemical Physics}}} \textbf{\bibinfo{volume}{4}}, \bibinfo{pages}{283--291}, \doiprefix\url{10.1063/1.1749836} (\bibinfo{year}{1936}).

\bibitem{landau1980statistical}
\bibinfo{author}{Landau, L.~D.} \& \bibinfo{author}{Lifshitz, E.~M.}
\newblock \emph{\bibinfo{title}{Statistical Physics: Volume 5}}, vol.~\bibinfo{volume}{5} (\bibinfo{publisher}{Pergamon}, \bibinfo{year}{1980}).

\bibitem{Hartmann:2005}
\bibinfo{author}{Hartmann, P.}, \bibinfo{author}{Kalman, G.~J.}, \bibinfo{author}{Donk\'o, Z.} \& \bibinfo{author}{Kutasi, K.}
\newblock \bibinfo{journal}{\bibinfo{title}{Equilibrium properties and phase diagram of two-dimensional yukawa systems}}.
\newblock {\emph{\JournalTitle{Phys. Rev. E}}} \textbf{\bibinfo{volume}{72}}, \bibinfo{pages}{026409}, \doiprefix\url{10.1103/PhysRevE.72.026409} (\bibinfo{year}{2005}).

\bibitem{LeBard:2012}
\bibinfo{author}{LeBard, D.~N.} \emph{et~al.}
\newblock \bibinfo{journal}{\bibinfo{title}{Self-assembly of coarse-grained ionic surfactants accelerated by graphics processing units}}.
\newblock {\emph{\JournalTitle{Soft Matter}}} \textbf{\bibinfo{volume}{8}}, \bibinfo{pages}{2385--2397} (\bibinfo{year}{2012}).

\bibitem{Fennell:2006}
\bibinfo{author}{Fennell, C.~J.} \& \bibinfo{author}{Gezelter, J.~D.}
\newblock \bibinfo{journal}{\bibinfo{title}{Is the ewald summation still necessary? pairwise alternatives to the accepted standard for long-range electrostatics}}.
\newblock {\emph{\JournalTitle{The Journal of Chemical Physics}}} \textbf{\bibinfo{volume}{124}}, \bibinfo{pages}{234104}, \doiprefix\url{10.1063/1.2206581} (\bibinfo{year}{2006}).

\bibitem{PhysRevResearch.5.013149}
\bibinfo{author}{Huang, D.}, \bibinfo{author}{Baggioli, M.}, \bibinfo{author}{Lu, S.}, \bibinfo{author}{Ma, Z.} \& \bibinfo{author}{Feng, Y.}
\newblock \bibinfo{journal}{\bibinfo{title}{Revealing the supercritical dynamics of dusty plasmas and their liquidlike to gaslike dynamical crossover}}.
\newblock {\emph{\JournalTitle{Phys. Rev. Res.}}} \textbf{\bibinfo{volume}{5}}, \bibinfo{pages}{013149}, \doiprefix\url{10.1103/PhysRevResearch.5.013149} (\bibinfo{year}{2023}).

\bibitem{PhysRevLett.82.5293}
\bibinfo{author}{Schweigert, I.~V.}, \bibinfo{author}{Schweigert, V.~A.} \& \bibinfo{author}{Peeters, F.~M.}
\newblock \bibinfo{journal}{\bibinfo{title}{Melting of the classical bilayer wigner crystal: Influence of lattice symmetry}}.
\newblock {\emph{\JournalTitle{Phys. Rev. Lett.}}} \textbf{\bibinfo{volume}{82}}, \bibinfo{pages}{5293--5296}, \doiprefix\url{10.1103/PhysRevLett.82.5293} (\bibinfo{year}{1999}).

\bibitem{PhysRevLett.42.795}
\bibinfo{author}{Grimes, C.~C.} \& \bibinfo{author}{Adams, G.}
\newblock \bibinfo{journal}{\bibinfo{title}{Evidence for a liquid-to-crystal phase transition in a classical, two-dimensional sheet of electrons}}.
\newblock {\emph{\JournalTitle{Phys. Rev. Lett.}}} \textbf{\bibinfo{volume}{42}}, \bibinfo{pages}{795--798}, \doiprefix\url{10.1103/PhysRevLett.42.795} (\bibinfo{year}{1979}).

\end{thebibliography}
\end{document}